\newcommand{\extendedversion}{final}
\documentclass[10pt,twocolumn,compsoc,journal,a4paper]{style/IEEEtran}

\usepackage[nocompress]{cite}

\input{documentProperties}

\begin{document}

\title{
Response-Time-Optimized Distributed \\
Cloud Resource Allocation
}
\date{}

\author{Matthias~Keller and~Holger~Karl%
\IEEEcompsocitemizethanks{%
\IEEEcompsocthanksitem Manuscript received January xx,yy; revised January xx, yy. This work was partially supported by the German Research Foundation (DFG) within the Collaborative Research Centre ``On-The-Fly Computing'' (SFB 901).
\IEEEcompsocthanksitem The authors are with the University
of Paderborn, Warburger Str. 100, 33098 Paderborn, Germany.
E-mail: mkeller@upb.de, holger.karl@upb.de%
}}

\IEEEtitleabstractindextext{%
\begin{abstract}
A current trend in networking and cloud computing is to provide compute resources widely distributed exemplified by initiatives like Network Function Virtualization.
This paves the way for a widespread service deployment and can improve service quality; a nearby server can reduce the user-perceived response times.
But always using the nearest server is a bad decision if that server is already highly utilized.

This paper investigates the optimal assignment of users to distributed resources -- a convex capacitated facility location problem with integrated queuing systems.
We determine the response times depending on the number of used resources. 
This enables service providers to balance between resource costs and the corresponding service quality.
We also present a linear problem reformulation showing small optimality gaps and faster solving times; this speed-up enables a swift reaction to demand changes.
Finally, we compare solutions by either considering or ignoring queuing systems and discuss the response time reduction by using the more complex model.
Our investigations are backed by large-scale numerical evaluations.

\end{abstract}
}

\maketitle

\IEEEdisplaynontitleabstractindextext

\IEEEpeerreviewmaketitle

\section{Introduction}
\label{sec:introduction}

\subsection{Challenges in Distributed Clouds}
\label{sec:chall-distr-cloud}

A current trend in networking and cloud computing is to provide compute resources widely distributed. Computation will not only take place on desktops or large data centres, but also at smaller centres or within the network itself, \eg, inside individual in-network server racks located near backbone routers. This trend is known under different labels, for example, Carrier Clouds~\cite{Taleb2014,Bagaa2014,Cai2013}, Distributed Cloud Computing~\cite{Endo2011,Agarwal2010,Pandey2010,Church2008}, or In-Network Clouds~\cite{Stolyar2013a,Scharf2012,Hao2009}. 
These In-Network Clouds tend to be less cost-efficient than conventional Clouds due to a worse economy of scale; they are hence often geared towards specific network services (\eg firewalls, load balancers).
Easing a more flexible deployment of these services became popular as Network Function Virtualization~\cite{Fischer2013} not only inside a data centre but also beyond, in wide area networks~\cite{Csaszar2013a,Taleb2013,Taleb2014}.
We consider not only executing network functions but more generically executing applications at those In-Network Clouds yielding an important advantage~\cite{Oberle2013,Alicherry2012}:
The resources of these Clouds are closer to end users than those of conventional Clouds, have smaller latency between user and cloud resource, and are therefore suitable for running highly interactive applications.
Examples for such applications are latency-critical applications~\cite{Wan2010, Barker2010}, user-customized streaming services~\cite{Bauer2011, Cucinotta2013, Ishii2011}, or Cloud Gaming~\cite{Lee2012}; the computing tasks range from processing the request, aggregating incoming data streams, up to rendering and encoding video streams.
In such applications, the crucial quality metric is the user-perceived \emph{response time} to a request as the application need to quickly react on user interactions.
Large response times impede usability, increase user frustration~\cite{Claypool2006, Lee2012}, or prevent commercial success.
An obvious solution to provide small response times would be to deploy an application at many sites so that each user finds one site nearby. 
This, however, is infeasible as each utilized site incurs additional costs. We are hence faced with the task to decide where a user's request shall be processed, using as few sites as possible at a best possible response time.  We refer to this task as the \emph{assignment problem}. This problem's trade-off between cost and quality is intuitive yet difficult to capture in a concrete problem statement and solution. 

\begin{figure}[t]
\centering
\begin{minipage}{40mm}
\includegraphics[width=\linewidth]{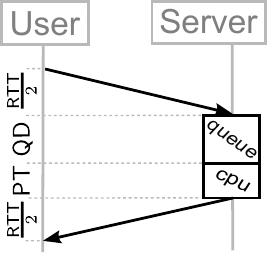}
\caption{Response time as the sum of round trip time~RTT, queuing delay~QD, processing time~PT.}
\label{fig:times_all}
\end{minipage}~\hspace{6mm}%
\begin{minipage}{40mm}
\vspace{3mm}
\includegraphics[width=\linewidth]{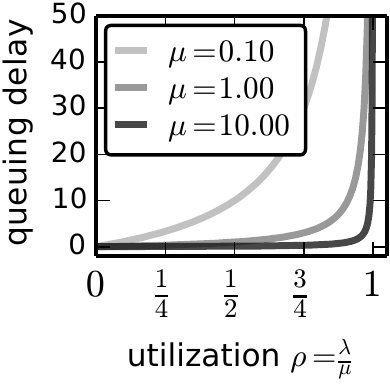}
\caption{Queuing delays QDs for different service rates.}
\label{fig:queuedelays}
\end{minipage}
\end{figure}

This difficulty lies in the nature of the response time. It is a sum of three parts: 
\begin{compactitem}
\item The network latency taken to send the request from the user to the cloud resource and sending the answer back -- the \emph{round trip time} (RTT); 
\item the actual \emph{processing time} (PT) of the request; 
\item the \emph{queuing delay} (QD) a request incurs at the cloud resource while other requests are currently processed at that resource (\Cref{fig:times_all}).
\end{compactitem}
In many applications, we can consider the processing time to be significantly smaller than the round trip time.
The round trip time depends on the choice where a user's request is processed and, to a much smaller degree, on the network load along the way. 
The queuing delay, however, depends on the sharing of a resource among many users and is not an effect immediately influenced by the decision for a single user; it depends on the joint decision for all users.

From queuing theory we know that, for a fixed utilization, the queuing delay is shorter for higher service rates.
\Cref{fig:queuedelays} shows queuing delays for different levels of system utilization $\rho{=}\nicefrac \lambda \mu$ for three different service rates $\mu$. 
For instance, web servers answering simple requests have high service rates and often negligible queuing delays (say, below \SI{10}{ms}).
This may explains why they are commonly ignored in literature (\Cref{sec:rw}).
We focus in this paper on computation-intensive applications, such as data processing, intrusion detection, or game render application (see next paragraph) oppose to light-weight application with nearly no computation like web servers just returning static content.
The processing time of computation-intensive applications is longer (\eg up to \SI{1}{\s}) than of light-weight applications, which implies a lower service rate and a longer queuing delay.
A long queuing delay becomes a large portion of the response time, significant large enough to necessitates considering them when deciding the assignments.

Shah\etalcite{Shah2015} survey intrusion detection systems and cite different measurements of packet processing times of up to \SI{10}{\ms}.  
Barker\etalcite{Barker2010} study game server map loads lasting $20$--$\SI{110}{\ms}$ in their experiments.
Ishii\etalcite{Ishii2011} conduct experiments on AWS~\cite{aws} using a parallel Data Processing Application and observe processing time between \num{400} and \SI{1800}{\ms}.
Lee\etalcite{Lee2012} and Claypool\etalcite{Claypool2006} observe drop in user experience when playing computer games with artificially increased latency to larger than $\SI{200}{\ms}$. 
In summary, we focus on applications with long processing times, $\SI{10}{\ms}$\,--\,$\SI{1}{\s}$, and on average internet round trip times, $\num{60}$--$\SI{600}{\ms}$~\cite{Kwon2014}.

Different proportions between round trip times and processing times are possible. 
Deciding the assignment in such scenarios can be simplified by ignoring the less dominant part:
Very short round trip times inside data centres, say less than \SI{20}{\ms}, leaves queuing delays as the dominant part of the response time.
Similarly, long processing times, say more than \SI{1}{\min}, renders queuing delays as the dominant part.
In both cases, round trip times can be ignored; doing so, the assignment problem becomes a simpler mapping problem.
On the opposite site, very short processing times, say \SI{0.1}{\ms}\,\footnotemark{}, result in very low queuing delays rendering the round trip time being the dominant part.
When ignoring queuing delays in this case, the assignment problem becomes a simpler, non-convex Facility Location Problem.
In summary, only if round trip times and processing times are not of the same magnitude, dropping the less dominant part is a vital option. 
If both times are of the same magnitude, also queuing delays have to be considered when deciding the assignment. 
Ignoring queuing delays in such cases worsen the assignment resulting in high response times; we compare both cases with different proportions of round trip times and processing times (\Cref{sec:eval:resptime}).

\footnotetext{Corresponding service rate of \SI{10000}{\req\per\second} of a high performance webserver, \eg Apache, Nginx, delivering static content.}

\begin{inextended}

\subsection{Queuing Delay Effects}
\label{sec:an-example}

\begin{figure}[t]
\centering
\includegraphics[width=45mm]{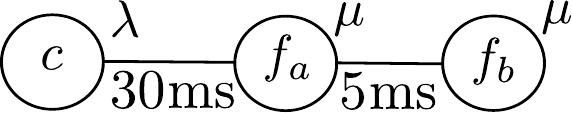}
\vspace*{6mm}
\includegraphics[width=9cm]{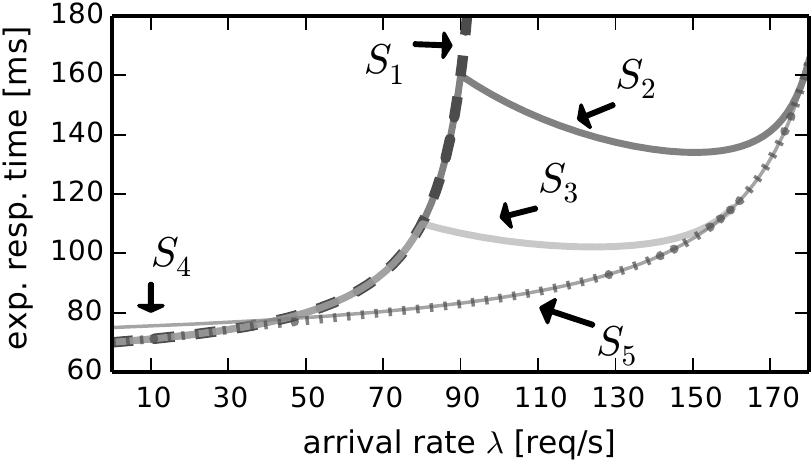}
\vspace*{-8mm}
\caption{The lower plot shows response times of different arrival rates~$\lambda_c$ for three different strategies for the topology. The upper plot shows the used topology.
\label{fig:example1}}
\end{figure}

How does the queuing delay affect the response time?
As a toy example, let us consider the network from \Cref{fig:example1} with three locations of interest:
One client $c$ and two possible facility locations $f_a$, $f_b$ with compute resources to run the application.
These resources are equally fast and can serve requests at rate $\mu{=}\SI{100}{\req\per\s}$.
Assume the round trip times between $c$ and $f_a$ as \SI{60}{\ms} and between $c$ and $f_b$ as \SI{70}{\ms}.
Requests enter the network at $c$ with arrival rate $\lambda$. 
With this setup, the requests can be served at only $f_a$, only $f_b$, or split among $f_a$ and $f_b$.
\Cref{fig:example1} shows, as a function of the arrival rate, the resulting response times RT = $\textnormal{RTT}$ + PT + QD for a few simple strategies $S_i$\footnotemark.

The first strategy $S_1$ minimizes only the requests' round trip time: 
Requests are assigned to the nearest facility $f_a$ and if its capacity $\mu$ is exceeded the remaining requests are assigned to $f_b$.
The dramatic response time growth for $\lambda{>}80$ is the result of too many requests assigned to facility $f_a$.
Let $\lambda_a$ be the assigned requests to $f_a$, then $f_a$'s utilization $\rho_a$ is $\nicefrac{\lambda_a}{\mu}$.
To avoid too large utilizations, strategies $S_2$ and $S_3$ limit them to a maximum value $\hat \rho$, $\rho_a,\,\rho_b \leq \hat \rho {<} 1$;
$S_2$ uses $\hat \rho {=} 0.9$ and $S_3$ uses $\hat \rho {=} 0.8$.
On the one hand $S_3$ with a lower limit has a shorter RT (\Cref{fig:example1}) than $S_2$ as the second facility is used earlier.
But on the other hand $S_2$ can handle a higher arrival rates than $ S_3 $,  $ 160{<}\lambda{<}180 $, because a higher limit enables handling more requests in total; system capacity is $\lambda {<} 2\mu \hat{\rho}$.
To relax such a predefined upper bound, $S_4$ dynamically adjusts the limit to the current system utilization, $\hat\rho {=} \nicefrac {\lambda} {2\mu}$.
With the same resources at both locations $S_4$ boils down to evenly splitting the load between the two facilities.
Compared to $S_{1..3}$ the resulting RTs are on the one hand small for $\lambda{>}40$ but on the other hand larger for $\lambda{<}40$.

So far, all assignment strategies ignore the resulting queuing delays.
In contrast, our last strategy $S_5$ additionally reduces the queuing delays of both resources.
The strategy's request assignment depends on the round trip times ($l_{cf}$) between both resources.
The resulting RTs are the lowest for all strategies $S_{1..5}$.
In conclusion, we were able to improve assignments by considering queuing delays.

Hereafter, we list the equations of the expected response times in \Cref{fig:example1}: $S_1$ to $S_3$ results in~\eqref{eq:example1a}, 
$S_4$ in~\eqref{eq:example1b}, and $S_5$ in~\eqref{eq:example1c}. 

\begin{IEEEeqnarray}{llr}
f_{\mu, \hat \rho} (\lambda) := & 
\begin{cases}
60 + \frac{1}{\mu - \lambda}, & \text{ if } \lambda {\leq}\hat \rho \mu   \\ 
\frac{\hat\rho\mu}{\lambda} \left( 60 + \frac{1}{\mu - \hat\rho\mu} \right) + \\
\quad \frac{\lambda - \hat\rho\mu}{\lambda} \left( 70 + \frac{1}{\mu - \lambda + \hat\rho\mu} \right),
& \text{ else }
\end{cases} \label{eq:example1a} \\
g_{\mu}(\lambda) := &  \frac 1 2 \left( 60 {+} \frac{1}{\mu {-} \nicefrac \lambda  2} \right) + 
\frac 1 2 \left( 70 {+} \frac{1}{\mu {-} \nicefrac \lambda 2} \right)
\label{eq:example1b} \\
h_{\mu}(\lambda) := &
\min_{\lambda_1 \in [0,\,\lambda]} \left \{ 
\begin{matrix}
\frac{\lambda_1}{\lambda} \left( 60 + \frac 1 {\mu - \lambda_1} \right) + \\
\frac{\lambda{-}\lambda_1}{\lambda} \left( 70 + \frac 1 {\mu - \lambda + \lambda_1} \right) \\
\end{matrix}
\right \} \label{eq:example1c} 
\end{IEEEeqnarray}

\begin{figure}[t]
\centering
\includegraphics[scale=1.0]{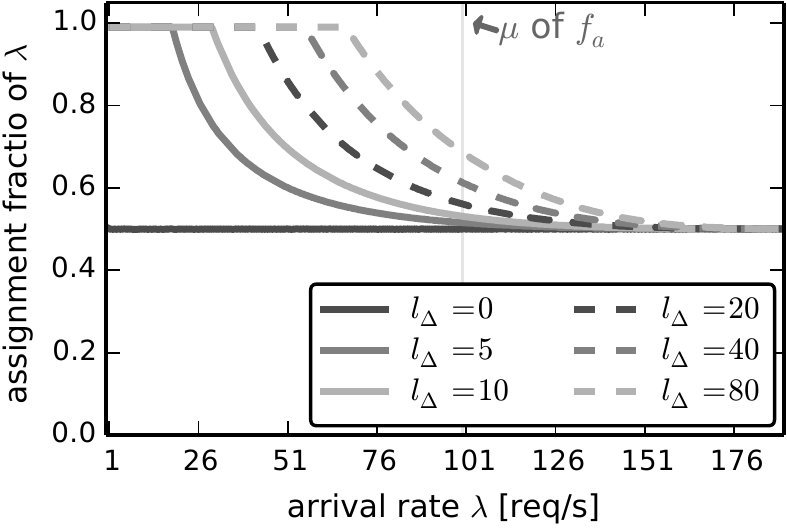} 
\caption{Request assignments to $f_a$ with different distances $l_\Delta$ between $f_a$ and $f_b$.}
\label{fig:diffweight}
\end{figure}

\Cref{fig:diffweight} shows strategy $S_5$'s request assignments to resource $f_a$ as a fraction of $\lambda$ on the vertical axis; the remaining requests are assigned to $f_b$. The horizontal axis shows an increasing arrival rate $\lambda$. The different lines correspond to distances $l_\Delta$ between $f_a$ and $f_b$ -- how much longer request transportation takes to send to $f_b$ instead to $f_a$. 
This way the original toy example is line $l_\Delta{=}10$ and the other lines vary the round trip time to $f_b$.
If the resources have the same round trip time, the assignment results in an even split. 
However, if one resource is farer away, $l_\Delta{>}0$, at first the nearer resource is preferred and only with increasing arrivals do the assignments converge to an even split. Then, the queuing delay portion of the response time is significantly larger than $l_\Delta$.

\end{inextended}

\subsection{Contribution}
\label{sec:contribution}
This paper discusses finding the optimal assignment between requesting users and compute resources hosting the answering application at different locations in the network. 
The assignment minimizes the expected average response time for all users.
We claim the necessity of considering request's queuing delays at used compute resources to avoid suboptimal assignments to \eg over-utilized resources (when round trip times and queuing delays are of the same magnitude).
To proof this claim, we present an extended Facility Location Problem with integrated queuing systems (\Cref{sec:convexopt}), show its convexity, and, for the first time, obtain optimal\footnotemark{} solutions for larger networks using a convex solver (\Cref{sec:convexopt}).
\footnotetext{\label{fn:opt}Numerically obtaining solutions by solver with a gap threshold of $10^{-6}$.}
Due to problem's complexity, solving times were already large for medium-sized networks (\eg \Cref{fig:deviation} \texttt{ta2} topology) hindering the large scaled evaluation we had envisioned.
We were able to shorten the solving times with high accuracy by non-trivially linearising the convex problem (\Cref{sec:linearapprox}). 
This linearised problem has a larger search space as the convex problem; despite this, it solves the original problem significant faster (empirically shown, \Cref{fig:deviation}).
Having now an adequate and fast substitution at hand enables us to compare numerous solutions obtained by considering and ignoring queuing delays supporting our claim by showing significant response time reduction when considering queuing delays (\Cref{sec:eval:qtornot}).
In addition, we show how the response time and queuing delays increases when using lesser compute resources (\eg fewer locations, \Cref{sec:eval:resptime}).

This evaluation in \Cref{sec:eval:qtornot} extends our own previous work~\cite{Keller2014}. 
We compare four factors influencing queuing delays and in addition vary input randomly in order to verify the statistical relevance of our findings. 
In summary, we solved and analysed \num{52500}~configurations.

\section{Related Work}
\label{sec:rw}
Assignment problems of the form described above have been investigated before. 
We structure their comparison along four dimensions relevant to this work:
Their model complexity, 
simplifications reducing the problem's search space,
optimization goals,
and solution approaches.
Finally, related systems of geographical load balancing are compared.

\subsection{Model Complexity}

The simplest model considers \emph{only the round trip time}~(RTT) when assigning users to cloud resources. They equate response time with RTT.
Clearly, this is a simplification of reality, yet minimizing this average RTT is equivalent to the well-known capacitated Facility Location Problem (FLP).
If the problem is further restricted to only use $p$ resources, it becomes a $p$-median FLP, which is NP-hard~\cite{Algorithmic1979}.

A step closer to reality is modelling \emph{also the processing time}~(PT) in addition to the RTT. But as long as PT is constant, this still stays a Facility Location Problem of the type described above. This can be easily seen by extending the original network topology by pseudo-links (at the server or user side) that represent these processing times via their latencies; this is a common rewriting technique for graph-based problems (including Facility Location Problems). 

The real challenge occurs when we also \emph{consider the queuing delay}~(QD).
In this case, the additional time cannot be expressed by rewriting the network topology as the QD depends on the assignment decisions: A higher utilization results in a longer wait, possibly trading off against a shorter RTT.

So far, this more general model has been considered only by few works discussed in the remaining of this section, most use simpler assumptions than ours~(\Cref{sec:problem}) rendering the problem easier to solve.
Vidyarthi\etalcite{Vidyarthi2009} allow the same degrees of freedom as we do.
They approximate, similar to us, the non-linear part of the objective function with a piece-wise linear function. 
However, in contrast to our work, they used a cutting plane technique which iteratively refines the piece-wise function as necessary; it remains unclear how large their linearisation error is.
In contrast, our evaluation shows small linearisation errors; and this is achieved by using a simpler technique.

\subsection{Simplifications}
Other authors investigate slightly different scenarios, so that their problem formulations are similar, yet simpler than ours.

Some authors~\cite{Zhang2012, Wang2002} replace the non-linear QD part with a constant upper bound and, consequently, 
the resulting problems become simpler to solve.
But this also hides QD changes as a result of assignment changes.
For instance, in a situation where load balancing would reduce the QD, this reduction is not visible as the QD part is constant.
Consequently, the resulting solution has further potential for optimization -- we exploit this potential.

In another simplification, the assignments are predefined by a rule. 
Some authors~\cite{Zhang2012, Aboolian2009, Wang2002} always assign requests to the nearest cloud resource.
In such a case, the problem reduces to just finding the best resource location and is easier to solve.
The assignments are then predetermined by the rule.
However, balancing the assignments could further reduce the QD but is not considered.
We do not use any predefined assignment rule, so we have the freedom to change assignments in order to further reduce the response times.

Another group of authors~\cite{Drezner2011, Berman2006} uses a parametrized assignment rule called the gravity rule: Weights determine how users are assigned to cloud resources.
These configurable weights are used to continuously solve the same problem with new weights reflecting the resource utilizations of the previous solution.
This approach does not guarantee to converge, so the authors propose a heuristic that attenuates the changes in each iteration, enforcing convergence with an unknown linearisation error.
In contrast, we solve the problem in one step by using all information to find the global optimum.

Liu\etalcite{Liu2015}, Lin\etalcite{Lin2012}, and Goudarzi\etalcite{Goudarzi2013} present a similar Facility Location Problem with convex costs such as queuing delays or resource's energy costs. 
In contrast to our work, they relax the integer allocation decision variable simplifying the problem to the cost of a less accurate solution when rounding up the obtained continuous allocations.
Our goal, in contrast, is to prevent unexpected expenses by introducing an upper bound to the number of used resources. 
Continuously relaxing our problem can cause any location to be allocated a bit and, consequently, any site is used and paid.
While the papers \cite{Liu2015,Lin2012} only consider queuing delays as a cost function, this paper discusses a holistic queuing system integration and additionally considers splitting and joining (assigning) of the arrival process.

\subsection{Optimization Goal}

Existing literature uses queuing delays in FLPs with three optimization goals: classic FLP, min/max FLP, and coverage FLP.

Classic FLPs are problems that minimize the average response time, like our problem (\Cref{sec:model}) or others~\cite{Wang2002,Berman2006,Vidyarthi2009,Pasandideh2010,Drezner2011,Zhang2012}, allowing RT variations for individual users. 

Aboolian\etalcite{Aboolian2009}'s min/max problem minimizes the maximum response time.
Intuitively, such problems improve especially the users' RT with high RTTs to cloud resources. 
However, if only one such user exists with resources being far away, assigning this user will negatively affect the assignments of other users:
Their assignments are now less-restrictedly constrained by a relaxed upper bound and are likely worse than without the first user.
In contrast, classical FLPs do not suffer this way from a worse case user.

Another type of problem is coverage problems; the user assignment's response times is upper bounded~\cite{Marianov1996,Marianov2002,Moghadas2011}. 
Structurally, a coverage problem is a special, simpler case of a min/max problem; 
the first has a predefined bound, which is additionally minimized in the second. 
Intuitively, such problems can be applied in scenarios where service guarantees for a certain maximal response time will be provided and paid.
In contrast, classical FLPs allow minimizing the average response time below the lowest possible response time bound.

\subsection{Solution Approaches}
A couple of heuristics were proposed solving related problems which are variants of the NP-hard capacitated FLP~\cite{Hakimi1979}. 
No work so far used solvers to obtain solutions (for non-relaxed problems) and full enumerations are known for small instances limited to open five facilities~\cite{Aboolian2009}.
A greedy \emph{dropping} heuristic successively removes from the set of candidates that resource which increases the response time by the smallest amount~\cite{Wang2002}. 
Greedy \emph{adding} heuristics successively add resources, which decreases the response time by the largest amount~\cite{Berman2006,Aboolian2009,Drezner2011}.
Another heuristic probabilistically selects set changes of used resources~\cite{Drezner2011} or performs a breath-first-search through ``neighbouring solutions'' where two solutions are neighbours if their sets of used resources differ in one element~\cite{Aboolian2009,Drezner2011}.
Such heuristics can be stocked in local optima and to mitigate this drawback meta-heuristics are used as a superstructure~\cite{Wang2002,Berman2006,Aboolian2009,Pasandideh2010,Drezner2011}.
These meta-heuristics typically refine previously generated initial solutions, which are obtained randomly or by combining existing solutions. 
The hope is that among the found local optima, one solution is very close to the global optimum -- but without any guarantee.
In contrast, we obtain global optima.
This is an important step for heuristic development as only this enables a clear judgement of heuristics' accuracy; their solution's gap to the global optimum.

Others~\cite{Wang2002,Vidyarthi2009} may achieve near optimal solutions by using optimization techniques like branch-and-bound and cutting planes but their solutions have unknown optimality gabs.
In summary, either optima for small input or solutions with unknown optimality gap are obtained.
This motivated our work on finding near-optimal solutions with a numerically very small optimality gap.

Liu\etalcite{Liu2015} and Wendell\etalcite{Wendell2010} present distributed algorithms for their global Geographical Load Balancing problem by decomposing it into separate subproblems solved by all clients. 
These subproblems converge to the optimal solution only if they are executed in several synchronized rounds in which assignment and utilization information are exchanged among all clients.
Both papers state that this distributed algorithms would obtain optimal solution faster than gathering everything to a centralised solver.
However, we believe that each round a communication delay is introduced when sending update information among all clients; they had ignored this delay in their evaluations. 
The resulting total delay over all rounds is likely larger than communicating with a centralised coordinator. 
In addition, our $p$-median Facility Location Problem has a global constraint on the maximal used resources preventing it to be easily separated into subproblems.

We observed that problem instances were solved only exemplary so far~\cite{Liu2015,Wendell2010,Lin2012,Marianov2002,Berman2006,Aboolian2009,Pasandideh2010,Pasandideh2010,Drezner2011}.
Consequently, the average performance of these solution approaches is hard to predict.
We go beyond this by undertaking a statistical performance evaluation.
We randomly vary our input data and verify the statistical relevance of our findings.

\subsection{Geographical Load Balancing}    

A system for Geographical Load Balancing (GLB) comprises two parts: The decision part selects appropriate server, sites, or Virtual Machines for requests of a certain origin -- the previous sections considers them.
This section focuses on the realisation part, which gathers monitoring information and implements selections.
Different middlewares had been proposed~\cite{Wendell2010,Freeman2006,Wong2006,Wong2005} which are shared between applications.
In this way, each application benefits from instances of the other application running at diverse sites by sharing monitoring information such as latency to servers or to customers.
They realise request assignments, \eg, to close-by or low utilised server by either configuring the Domain Name System (DNS) or are explicitly queried ahead a request send.
Slightly different, Cardellini\etalcite{Cardellini2000} propose redirecting requests to different sites to balance the load. Policies ranges from redirecting all, only largest, or only group requests to selecting sites based on round-robin, site utilization, or connection properties.
Our paper focuses on solving the problem and investigates whether the complexer problem with queuing systems is worth the additional efforts and our results can be applied to improve geographical load balancing systems.

\section{Problem}
\label{sec:problem}
This section first formalises our scenario model and then details on practical realisations.
Afterwards this section discusses problem's convexity and proposes a problem linearisation minimizing the maximal linearisation error.

\subsection{Model}\label{sec:model}

\begin{figure}[t]
\centering
\includegraphics[width=9cm]{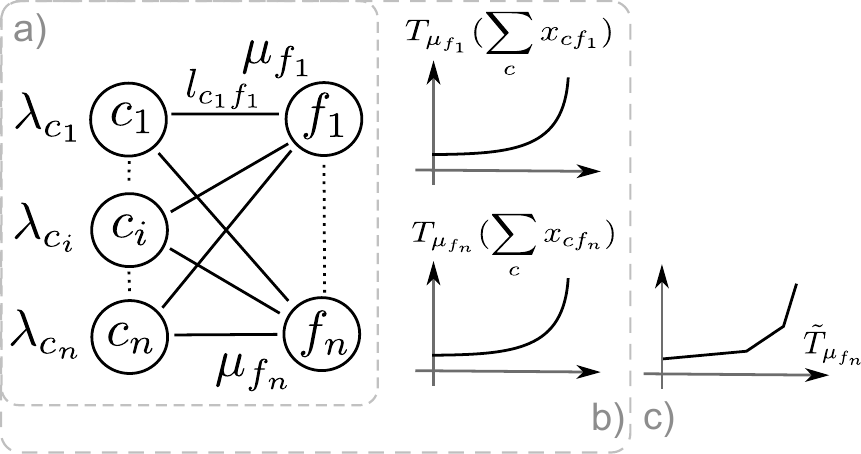}
\caption{Bipartite graph of a Facility Location Problem~(a); time-in-system functions at each facility~(b) and, alternatively, piece-wise linearised functions~(c).}
\label{fig:problemvis}
\end{figure}

\begin{table}[t]
\caption{Model variables} \label{tbl:var} %
\vspace{-3mm}
\centering
\begin{tabular}{p{20mm}p{52mm}}
\hline 
\multicolumn{2}{l}{Input constants:} \\
\hline
$G=(V,E)$ & Bipartite graph with $V=C \cup F$, {$ C \cap F = \varnothing$}  with client nodes $c \in C$ and facility nodes $f \in F$ \\
$\lcf \in \mathbb{R}_{>0}$ & Round trip time between $c$ and $f$  \\ 
$\mu_f \in \mathbb{R}_{>0}$  & Service rate as capacity at $f$ \\ 
$\lambda_c \in \mathbb{R}_{>0}$   & Arrival rate as demand at $c$ \\ 
$\funcT \in \mathbb{R}_{>0}$ & Time in queuing system (TiS) \\
$\alpha_{\mu s}, \beta_{\mu s}$ & $s$-th basepoint $\funcT(\alpha_s) {=} \beta_s$ of $\funcTa$ \\[1mm]
\multicolumn{2}{l}{Decision variables:} \\
\hline 
$\xcf \in \mathbb{R}_{>0}$  & Assignment in demand units \\ 
$y_f \in \{0,1\} $ & Indicator if $f$ is opened ($=1$) \\ 
$\zfs \in [0,1] $ & Weight of $s$-th basepoint at $f$ \\[1mm]
\multicolumn{2}{l}{Helper variables:} \\
\hline 
$\Lambda$; $\Lambda_f$ & total $\Lambda{=}\sumc \lambda_c$; at $f$: $\Lambda_f {=} \sumc \xcf$ \\
$\tau \in \mathbb{R}_{>0}$ & Sufficient small value \\
$\rho \in \mathbb{R}_{>0} {=} \nicefrac \lambda \mu$ & System utilization \\
\hline 
\end{tabular} 
\end{table}

Our scenario is formalized as a capacitated $p$-median Facility Location Problem~\cite{Drezner2004}.
A bipartite graph $G = (C {\cup} F,\, E)$ has two types of nodes: \emph{clients} ($c {\in }C$) and \emph{facilities} ($f {\in} F$). 
Clients correspond to locations where user request flows enter the network.
Facilities represent candidate locations to execute the application, \eg, data centres.
More precisely, a (compute) resource refer to a host at such a data centre executing the application.
\Cref{fig:problemvis}a shows such a graph.
The geographically\footnotemark\ distributed demand is modelled by the request arrival rate $\lambda_c$ for each client $c$.
\footnotetext{More precisely, the request arrival and service points are topologically distributed; the round trip time of a path between two points only roughly matches its geographically distance. We use ``geographically'' for a convenient explanation.}
Computing capacity is modelled as the request serving rate $\mu_f$ for each facility $f$.
The round trip time $l_{cf}$ is the time to send data from $c$ to $f$ and back.
\Cref{tbl:var} lists all variables.

Our first problem formulation recapitulates the known $p$-median problem $\optP(G,\, \lambda,\, \mu,\, p)$:~
\begin{IEEEeqnarray}{lllr}
\min_{x} \quad & \IEEEeqnarraymulticol{2}{l}{\frac{1}{\sum_c \lambda_c} \sumc \sumf \xcf \lcf} & \textnormal{(objective)} \label{eq:p:obj} \\
\textnormal{s.t.} 
&    \sumf \xcf = \lambda_c, \qquad & \forall c \qquad & \textnormal{(demand)} \label{eq:p:demand}\\
&    \sumc \xcf \leq y_f \mu_f, \qquad & \forall f & \textnormal{(capacity)} \label{eq:p:capacity} \\
&    \sumf y_f = p & & \textnormal{(limit)} \label{eq:p:limit}
 \end{IEEEeqnarray}

The formulation contains two decision variables: $x_{cf} {\in} \mathbb{R}_{\geq 0}$ describes which part of $c$'s request rate $\lambda_c$ is assigned to which $f$; $y_f {\in} \left\lbrace 0, \, 1  \right\rbrace$ describes if location $f$ is used or not.
The objective is to minimize the average response time; but without modelling the queuing delay and service time at facilities, the response time only consists of the round trip time. 
The RTT is minimized while all demand is served (\ref{eq:p:demand}) and the capacity is not exceeded (\ref{eq:p:capacity}).

In addition, exactly $p$~locations are used (\ref{eq:p:limit}).
This constraint serves two proposes. 
First, by limiting the number of location where the application is developed to, the expenses for the application provider when leasing Cloud resources is bound.
In Facility Location variants where facility opening costs are directly integrated the resulting total costs are unsure.
Second, stating the problem with this bound allow us to investigate the response time trend while allowing more and more resources (\Cref{sec:eval:resptime}).
Since 1979 the problem without capacity is known to be NP-hard~\cite{Hakimi1979}. 
This problem is a generalization and, thus, also NP-hard.

Until now, the response time has only been the round trip time. 
To predict the queuing times, the model is extended by queuing systems at each facility~(\Cref{fig:problemvis}b).
There, the service times are exponential distributed.
The inter-arrival times at each node $c$ are described by a Poisson process.
The requests can be assigned to multiple facility ($\sum_f x_{cf}$) and, there, the individual assignment from different nodes are aggregated ($\sum_c x_{cf}$).
The resulting process is also a Poisson process, because splitting and joining does not change the underlying random distribution. 
As a result, we have a M/M/1-queuing model~\cite{Bolch2005}.
The function for the time in queuing system (TiS) computes the processing time plus the queuing delay (\Cref{fig:times_all}), $\funcT(\lambda) {=} \frac{1}{\mu-\lambda}$. 
Putting everything together, the corresponding formulation of this queuing-extended $p$-median problem $\optQP(G,\,\lambda,\, \mu,\, p)$ is:~%
\begin{IEEEeqnarray}{lllr}
\min_{x,\,y} \quad & \IEEEeqnarraymulticol{3}{l}{
    \underbrace{ \frac{\sum_{cf} x_{cf} l_{cf} }{\sum_c \lambda_c}  }_{\textnormal{average RTT}} +
    \underbrace{ \frac{
       \sum_{f} \left( \sum_c x_{cf} \right ) 
        \frac{1}{ \mu_f{-}\sum_c x_{cf}}
       }{\sum_c \lambda_c}}_{\textnormal{average TiS}} 
}\label{eq:qp:obj}   \\
\textnormal{s.t.} 
&    \sumf \xcf = \lambda_c, \qquad & \forall c \quad & \textnormal{(demand)} \qquad \label{eq:qp:demand}\\
&    \sumc \xcf < y_f \mu_f, \quad & \forall f & \textnormal{(capacity)}  \qquad \label{eq:qp:capacity} \\
&    \sumf y_f = p & & \textnormal{(limit)} \qquad \label{eq:qp:limit}
 \end{IEEEeqnarray}

The new objective (\ref{eq:qp:obj}) is to minimize the average response time, which is the sum of the average round trip time and the average time in system (\Cref{fig:times_all}). 
\Coref{eq:qp:demand} is the same as \Coref{eq:p:demand}; all demand must be served.
\Coref{eq:qp:capacity} assures the \emph{steady state} ($\lambda {<} \mu$, c.f.~\cite{Bolch2005}) for each queuing system. 
Finally, \Coref{eq:qp:limit} mandates to use exactly $p$ locations, just like \Coref{eq:p:limit}.

\subsection{System design} \label{sec:sysdesign}
The presented optimisation problem $\optQP$ is part of a large system which dispatches requests of a certain origin to sites as decided.
Examples of such a system ranges from Geographical Load Balancing systems (\Cref{sec:rw}) to our own Application Deployment Toolkit~\cite{Keller2013b}.
They monitor traffic, decide assignments, and reconfigure the dispatching subsystem in time periods.
The average arrival rate~$\lambda_c$ is the averaged number of incoming requests at router~$c$ for the last period. 
By solving problem $\optQP$ once a period, the request assignments\footnotemark{} are decided for the next period.
The decision is realised by configuring the dispatching subsystem, \eg, DNS, and allocating cloud resources accordingly.
\footnotetext{The assignment $x_{cf}$ is the \emph{request rate} dispatched from $c$ to $f$, in short \emph{request assignment}.}
The system is not meant to allow a fine grained assignment decision for each incoming request, \eg at line speed.
On longer terms, it decides which sites are strategically used and how incoming requests are roughly distributed.

\subsection{Convex Optimization} \label{sec:convexopt}

Previous work (\Cref{sec:rw}) also considered our objective function (\ref{eq:qp:obj}) but did not solve the corresponding problem optimally, except for small graphs via full enumeration.
This is because of the non-linearity of the objective function which necessitates non-linear solvers.
There exist a couple of non-linear solvers with different specializations: quadratic, convex, or non-convex objective functions.
By determining the complexity class of our objective function, we can choose a suitable solver, to efficiently obtain a global optimum.

This section first proves the objective function's convexity and shows that it is not simpler, \eg, quadratic. 
Afterwards, it describes how we used the convex solver.

\begin{definition} \label{def:convex}
A function $g$ is convex if its domain $\dom(g)$ is a convex set and if $g''(x) {\geq} 0$ holds $\forall x {\in} \dom(g)$~\cite{Boyd2004}. 
\end{definition} %

\begin{lemma} \label{def:convexsum2}
Function $g {=}\sum_i w_i g_i$, $g{:}\mathbb{R}^n {\rightarrow} \mathbb{R}$, is convex, iff $\forall i{:}\; g_i{:}\mathbb{R}^n {\rightarrow} \mathbb{R}$ and $w_i {\in} \mathbb{R}_{>0}$ is convex \cite{Boyd2004}.
\end{lemma} %

\begin{theorem}\label{thm:convexopt}
The objective function (\ref{eq:qp:obj}) of $\optQP$ is convex with function $T_\mu$ computing the sojourn time in an \qmmone queuing system. 
\end{theorem}

\begin{IEEEproof}
The domain of $\funcT(\lambda){=}\nicefrac{1}{\mu-\lambda}$ is the interval $0 {\leq} \lambda {<} \mu$ enforced by constraint~(\ref{eq:qp:capacity}); an interval is always a convex set. 
The second derivative $\funcT''(\lambda){=}\nicefrac{2}{ (\mu{-}\lambda)^3}$ is always larger $0$ within its domain. 
By \Cref{def:convex}, $\funcT(\lambda)$ is a convex function.

For a fixed $f$ in the objective function, $0 {<} \Lambda_f  {=} \lambda {<} \mu$ and $\funcT(\Lambda_f)$ is convex. Then, the non-negative weighted sum of convex functions $\sum_f \Lambda_f \funcT(\Lambda_f),\, \Lambda_f {=} \sumc \xcf$ is also convex (\Cref{def:convexsum2}). The term remains convex after $\nicefrac{1}{\Lambda} {>} 0$ is multiplied.
The left term of the objective function is linear and also convex.
Since the sum of two convex functions is convex, the objective function (\ref{eq:qp:obj}) is convex.
\end{IEEEproof}

With the knowledge of a convex objective function, we can ignore less efficient solvers for more general, non-convex problems.
The next more efficient solver class is quadratic, which need objective functions of the form $x^TMx$ with symmetric matrix $M {\in} \mathbb{R}^n$. 
But our objective function is not of this form, making quadratic solvers inapplicable.
Consequently, we have to use a convex solver.

\textbf{Implementation:}
We choose the optimization framework \mbox{CVXOPT}~\cite{cvxopt} from the authors of \cite{Boyd2004}.
$\optQP$ is a mixed integer problem, which is not directly supported by \mbox{CVXOPT}.
Continuously relaxing the problem is not possible (\Cref{sec:rw}).
We decomposed $\optQP$ into solving multiple subsets $F'$ of $F$ with $|F'|=p$:
\begin{IEEEeqnarray}{lr}
\optQP(G=(C \cup F, E), \, \lambda, \, \mu,\,p ,\, \tau) =   \notag \\
\hspace*{1em} \min_{F'{\in}F,\,|F'|{=}p} \left\lbrace \; 
            \optPQP((C \cup F', E), \, \lambda, \, \mu,\, \tau)  \; \right\rbrace \qquad 
\end{IEEEeqnarray}
with the purely convex sub-problem $\optPQP(G ,\, \lambda_c ,\, \mu_f ,\, \tau)$:
\begin{IEEEeqnarray}{lllr}
\min_x \quad & \IEEEeqnarraymulticol{3}{l}{
    \frac{\sum_{cf} \xcf \lcf}{\sum_c \lambda_c}  + 
    \frac{\sumf \frac{\sum_c \xcf}{\mu_f {-} \sum_c \xcf} }{\sum_c \lambda_c}  
}  \label{eq:pqp:obj} \\
\textnormal{s.t.} 
&    \sumf \xcf = \lambda_c, \qquad & \forall c \qquad & \textnormal{(demand)} \label{eq:pqp:demand}\\
&    \sumc \xcf \leq \mu_f - \tau, \quad & \forall f & \textnormal{(capacity)} \label{eq:pqp:capacity} 
 \end{IEEEeqnarray}

The decomposition optimally solves $\optQP$ by solving a non-integer convex subproblem~$\optPQP$ several times for different configurations of the binary variables~$y_q$; these variables indicate which facility is used.
First, problem~$\optPQP$ is solved with all facility subsets $Q'{\subseteq} Q$ with $|Q'|{=}t$.
Then, one of all problem~$\optPQP$'s solution is selected that has the minimal response time~\eqref{eq:pqp:obj}.
In this solution, the decision vector~$x$ equals the $\optQP$'s decision vector~$x$ and problem~$\optQP$'s decision vector~$y$ is represented by subset $Q'$, $\forall q{\in}Q'{:}\;y_q{=}1$.
In this way, the optimal solution for problem~$\optQP$ is found.

\mbox{CVXOPT} solves $\optPQP$ by checking the domain (constraints) and iterating towards the optimum by using the Jacobi and Hessian matrix (first and second order derivatives) of the objective function~(\ref{eq:pqp:obj}). 
Hardcoding such matrices is not feasible for a large number of parameter configurations.
We want to have an automated solution obtaining these matrices at runtime. 
Algebra systems like Maxima\footnote{Maxima manuel: \url{http://maxima.sourceforge.net/docs/manual/maxima.pdf}} can be used, but need a detour through another system and computing derivatives of multi-dimensional functions takes time;  
for small input, more time than solving the problem.
To obtain these matrices faster, we found, not too surprisingly, that the structure of (\ref{eq:pqp:obj}) and its derivatives are the same for different $|C|,\,|F|$.

Exploiting this property, we were able to deduce a construction rule\ifextended[\footnotemark{}]{} for both matrices.
Using this rule, we construct our Jacobi and Hessian matrices at runtime for different inputs without notable overhead.
\ifextended[\footnotetext{This paper's extended version deduces the construction rules.}]{}

\begin{inextended}
In detail, we constructed the Jacobi and Hessian matrices from the objective function~(\ref{eq:pqp:obj}); here, reintroduced as a convenient copy $f$ (\ref{eq:mat:f}), $f(x_{11},...,x_{cf})$.
{
\begin{IEEEeqnarray}{llr}
f(x)= &\frac{1}{\Lambda} \sum_{cf} l_{cf} x_{cf} + \frac{1}{\Lambda} \sum_f \frac {\Lambda_f}{\Lambda_f{-}\mu_f}, \notag \\
    & \qquad \Lambda = \sum_c \lambda_c, \, \forall f{:}\; \Lambda_f = \sum_c x_{cf} \label{eq:mat:f} 
\end{IEEEeqnarray}
}
The Jacobi matrix~\eqref{eq:mat:d1} for one function is a vector of partial derivatives $(a_{11},...,a_{cf})$ for each variable $\xcf$.
{\small
\begin{IEEEeqnarray}{llr}
J_f(x)= &\left( a_{cf} \right)_{cf \in C \times F}, \,  \notag \\
    & a_{cf} = \frac {\Lambda_f} {\Lambda (\mu_f {-}\Lambda_f)^2} {+}  \frac {1}{\Lambda (\mu_f{-}\Lambda_f)} + \sumc \frac {l_{cf} }{\Lambda} \label{eq:mat:d1} \qquad 
\end{IEEEeqnarray}
}
Structurally, $f(x)$ is a sum of terms, and differentiating $f(x)$ can be done by differentiating the terms individually and afterwards summing all terms up. 
Two types of terms exist (\ref{eq:mat:g11}) with different derivatives (\ref{eq:mat:g2}, \ref{eq:mat:g1}). 
In Jacobi matrix ($\ref{eq:mat:d1}$), each partial derivative $a_{cf}$ is $g'_{1,cf}(x) + g'_{2,cf}(x)$.
\begin{IEEEeqnarray}{lr}
g_{1}(x) = \frac {l_{cf}x_{cf}}{\Lambda} \; g_{2}(x) = & \frac {\Lambda_f}{\Lambda (\Lambda_f{-}\mu_f)} \label{eq:mat:g11}   \qquad \end{IEEEeqnarray}%
\begin{IEEEeqnarray}{lllr}
g'_{1,cf}(x) & = 
\frac {d} {dx_{cf}}\frac {l_{ij}x_{ij}}{\Lambda} = 
    \begin{cases}
    \frac {l_{ij}}{\Lambda} & \text{ if } cf{=}ij\\ 
    0 & \text{ else }    
    \end{cases} \label{eq:mat:g2}   \qquad \\ 
g'_{2,cf}(x) & = 
\frac {d} {dx_{cf}}\frac {\Lambda_j}{\Lambda (\Lambda_j{-}\mu_j)}  \notag \\
& = 
    \begin{cases}
    \frac {\Lambda_j} {\Lambda (\mu_j -\Lambda_j)^2} + \frac {1} {\Lambda_j (\mu_j - \Lambda_j)} & \text{ if } f{=}j\\ 
    0 & \text{ else }    
    \end{cases} \label{eq:mat:g1} \qquad   
\end{IEEEeqnarray}
Similarly, the Hessian matrix~\eqref{eq:mat:d2} contains second-order, partial derivatives which are first derived in $\xcf$ direction (rows) and then in $x_{de}$ direction (columns).
{
\begin{IEEEeqnarray}{llr}
H_f(x) = &\left( a_{cf\,de} \right)_{cf\in C \times F,\, de \in C \times F}, \, \notag \\
    & a_{cf\,de} = 
    \begin{cases}
     \frac {2 \Lambda_f} {\Lambda (\Lambda_f {-} \mu_f)^3} + \\
     \qquad  \frac {2} {\Lambda (\Lambda_f {}- \mu_f)^2} & \text{ if } f{=}e\\ 
    0 & \text{ else }
    \end{cases} \label{eq:mat:d2}
\end{IEEEeqnarray}
}
Each cell $a_{cf\,de}$~(\ref{eq:mat:d2}) is $g''_{1,cf\,de}(x) + g''_{2,cf\,de}(x)$ from~(\ref{eq:mat:g3}, \ref{eq:mat:g4}).
\begin{IEEEeqnarray}{lllr}
g''_{1,cf\,de}(x) =& 
\frac {d} {dx_{cf}\,dx_{de}}\frac {l_{ij}x_{ij}}{\Lambda} &=  0 \label{eq:mat:g3}   \qquad \\ 
g''_{2,cf\,de}(x) =& 
\frac {d} {dx_{cf}\,dx_{de}}\frac {\Lambda_j}{\Lambda (\Lambda_j{-}\mu_j)} &    \notag \\
\IEEEeqnarraymulticol{3}{l}{
\qquad = 
    \begin{cases}
    \frac {2 \Lambda_j} {\Lambda (\Lambda_j - \mu_j)^3} +  \frac {2} {\Lambda (\Lambda_j - \mu_j)^2} & \text{ if } f{=}j \wedge e {=} j\\ 
    0 & \text{ else }    
    \end{cases} \label{eq:mat:g4} \qquad 
}
\end{IEEEeqnarray}

\end{inextended}

\subsection{Linear Approximation} \label{sec:linearapprox}
While \mbox{CVXOPT} solves the problem optimally, it has to test all subsets $F'$, which takes time. 
As an alternative, the convex objective function is linearised.
This way, well researched linear solvers can be used to obtain solutions faster.

\subsubsection{Piece-wise linear}\label{sec:linpwl}
Any non-linear function~$g(x) {:} \mathbb{R} {\rightarrow} \mathbb{R}$ over a finite interval $[\alpha_0,\,\alpha_{m-1}] {\subset} \mathbb{R}$ can be approximated by a piece-wise linear (PWL) function $\tilde{g}$~\cite{MINLP2012}. 
This function consists of $m$ basepoints {$\alpha_0,..,\,\alpha_s,..,\,\alpha_{m{-}1}$}, corresponding function values {$\beta_s {= }g(\alpha_s)$}, and is defined in (\ref{eq:defpwl}) for $\alpha_s {\leq} x {\leq} \alpha_{s+1}$. 
\begin{IEEEeqnarray}{llr}
\tilde{g}(x) & :=
(x-\alpha_s) \frac{(\beta_{s+1}-\beta_s)}{(\alpha_{s+1}-\alpha_{s})} + \beta_s, \notag \\
& \alpha_s  \leq x \leq \alpha_{s+1} \quad \forall s {\in} [0, m{-}2]
& \quad \label{eq:defpwl}
\end{IEEEeqnarray}

\begin{figure}[t]
\centering
\begin{inextended}
\includegraphics[width=9cm]{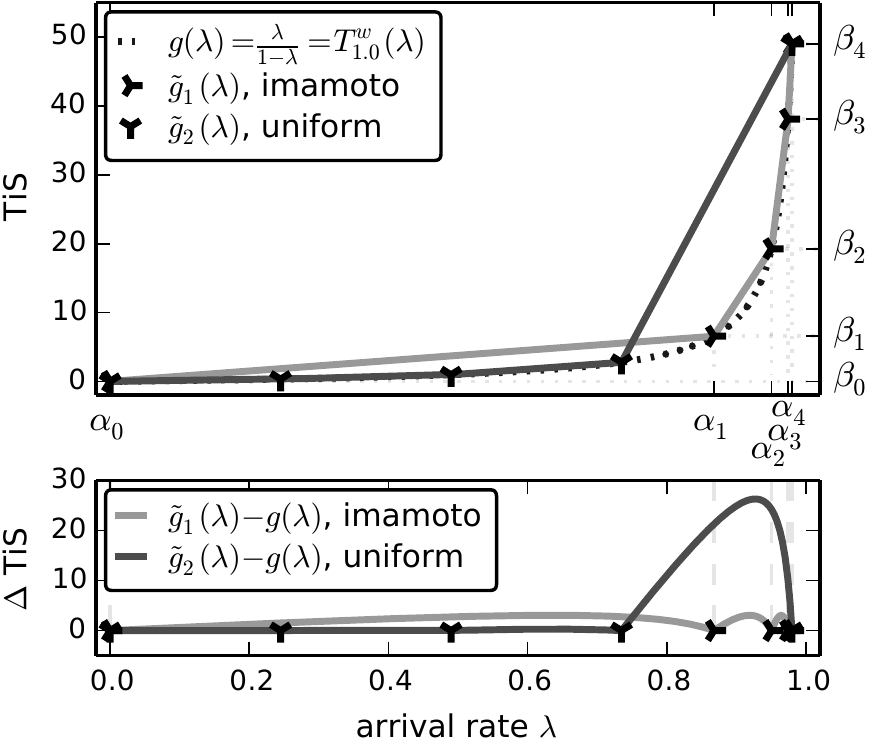}
\end{inextended}
\begin{notinextended}
\includegraphics[width=7cm]{./work/compare_pwlerror2}
\end{notinextended}
\caption{
The top plot shows an example function $g$ with two possible linearizations with the same number of basepoints.
The bottom plot shows absolute differences between the linearizations and $g$. 
Imamoto's linearisation has a smaller maximum difference.}
\label{fig:pwl}
\end{figure}

As an example, let us consider the part $\lambda \funcT(\lambda)$ of~\eqref{eq:qp:obj} for~$\mu{=}1.0$.
Then $g(\rho) {:=} \lambda \textnormal{T}_{1}(\lambda) {=} \nicefrac \lambda {1-\lambda}$ is our example function to linearise.
\Cref{fig:pwl}a shows $g$ and two different linearisations $\tilde g_1$ and $\tilde g_2$.
The horizontal axis shows the arrival rate and the vertical axis shows the corresponding TiS.
\Cref{fig:pwl}b shows the absolute differences between $g$ and either linearisation $\tilde g_1$ or $\tilde g_2$.
These differences denote the linearisation accuracy: 
The smaller the differences are, the tighter the PWL function resembles the original function.
We use the maximum of all absolute differences $\epsilon_{\tilde g}$, defined in~\eqref{eq:defmaxerr}, to measure the \textit{linearisation accuracy}. 
We seek basepoints $\alpha_i$ that minimize this error.
\begin{IEEEeqnarray}{llr}
\epsilon_{\tilde g} & := \max_{x \in [\alpha_0,\,\alpha_{m-1}]} \left| \tilde g(x) - g(x) \right| & \label{eq:defmaxerr}
\end{IEEEeqnarray}

Given a set of basepoints resulting in a certain error, this error is reduced by placing an additional basepoint at a point where the absolute difference equals the error. 
However, more basepoints also increase the number of necessary variables for the optimization, which increases search space and solving runtime.

Some functions are hard to approximate with linear segments, \eg, functions with large second-order derivative values.
If their values are large within the linearisation interval $[\alpha_0,\,\alpha_{m-1}]$, the error will be large. 
The TiS function's asymptote $\lim_{\lambda \rightarrow \mu} \funcT(\lambda){=}\infty$ approximated by linear segments results in such a larger error. 
One possible control knob is to adjust the interval $\alpha_{m-1} {<} \mu$.
But this also introduces an artificial capacity limit: Small values (\eg $\alpha_{m-1}{=}0.8\mu$) result in fewer requests served than possible (\cf \Cref{fig:example1}b's $S_2$).
Consequently, the total arrival rate for which solutions are feasible to obtain is smaller, $\nicefrac {\sum_c  \lambda_c}{p}  {\leq} \alpha_{m-1} {<} \mu$ with $p$ used resource. 

Both PWL functions in \Cref{fig:pwl}b, uniform and imamoto, have the same number of basepoints but at different positions.
As shown, uniformly distributing the basepoints can dramatically increase the error~($\tilde g_1$).
In contrast, the grey basepoints have small errors~($\tilde g_2$). 
Those basepoints were computed by our algorithms detailed in \Cref{sec:linalgo}.

We evaluate the first two control knobs, the number of basepoints and the linearisation interval's upper bound, in \Cref{sec:eval:lin}. 
For the third control knob, the basepoint positions, our algorithm determines basepoints with low error.

\newcommand{\varaneu}{\alpha_s^\textnormal{neu}}
\newcommand{\varaalt}{\alpha_s^\textnormal{alt}}
\subsubsection{Linearisation algorithm}\label{sec:linalgo}
Our algorithm obtains basepoints for convex functions with low error. 
It is an extended version of Imamoto's~algorithm~\cite{Imamoto2008}.
Imamoto's~algorithm iteratively refines $m$ basepoints by moving them individually along the abscissa to reduce the error $\epsilon_{\tilde g}$.
Each basepoint's adjustment $\Delta_s$ along the abscissa, $\varaneu {=} \varaalt {+} \Delta_s$, is computed from the basepoint's first-order derivative $\frac {d}{d\varaalt}g(\varaalt)$ and the inter-basepoint distance $d_s {=} \alpha_s {-} \alpha_{s{+}1}$.

The paper's~\cite{Imamoto2008} statement is that the algorithm computes basepoints which have the maximal linearisation accuracy for the given number of used basepoints.
However, the algorithm runs in numerical issues rendering the algorithm useless for some convex functions.
When fixing\footnotemark\ them, it cannot be guaranteed any more that the resulting basepoints form an linearisation with maximal accuracy (minimal error). 
But it is still very small -- still a good and fast option to linearise convex functions.
\footnotetext{This paper's extended version details our improvements of Imamoto's~algorithm.}

\begin{inextended}
More in detail, we extend  Imamoto's~algorithm~\cite{Imamoto2008} and fixed the following two cases:
First, the algorithm iteratively adjusts the current set of basepoints so that the error  is successively reduced.
These adjustments are weighted in order to allow gradually finer changes so that the error after each iteration converges to the minimum error in theory.
In practice, floating-point accuracy is limited and sometimes values are too small, changes not applied, and the algorithm iterates infinitely.
We fixed that by additionally aborting if no further basepoint changes are observed.

Second, for special functions the algorithm terminates with a division by zero.
The cause is computing a basepoint $\alpha_s$'s adjustment $\Delta_s$ depending on original function's derivative
$g'(\alpha_s) {=} \frac {d} {d\alpha_s} g(\alpha_s)$.
The division by zero occurs if the difference of two values $g'(\alpha_i)$ numerically equals zero, 
$\exists i{\not=}s : g'(\alpha_s){-}g'(\alpha_i) = 0$.
That is if $g$ resembles a linear function over some interval.
We fixed that by removing all basepoints $\alpha_s$ with 
$g'(\alpha_s){=}g'(\alpha_i)$, $i{<}s$ and inserting these basepoints between basepoints whose $g'(\alpha)$ values differ from each other. 
This assures that the error never increases or can be reduced: 
For those intervals of the function which are nearly linear, a linearisation over the whole interval yields a low error; thus, removing basepoints within this interval has little impact. 
Inserting these basepoints at another non-linear part of the function improves the linearisation accuracy as the PWL function becomes tighter.
\end{inextended}

\subsubsection{Formulation of linearised problem} 
This section describes the problem reformulation using a PWL function.
From existing alternatives~\cite{Padberg2000}, we used a \emph{Special Ordered Set} (SOS$k$) of type $k=2$ (SOS$2$)~\cite{Beale1970}: 
In a set of continuous variables, at most $k$ of them, adjacent to each other, may take non-zero values. 
Current linear solvers directly support SOS$2$.

A PWL function $\tilde y {=} \tilde g(x)$ is represented by a set of $m$ continuous decision variables {$0 {\leq} z_s {\leq} 1$} with a SOS$2(z_0,..\,z_s,..\,z_{m-1})$ constraint and a convex combination $1 {=} \sums z_s$.
This way, two adjacent values sum up to $1 = z_s {+} z_{s{+}1}$. 
These values are then used as weights for the basepoints $(\alpha_s,\,\beta_s)$ obtained previously by the linearisation process (\Cref{sec:linpwl}). 
This way, the weighted sum of all basepoints results in the piece-wise linear problem, $x {=} \sums z_s \alpha_s $, $y {=} \sums z_s \beta_s$.

Using this representation, we linearise the convex part of the objective function (\ref{eq:qp:obj}) $\Lambda_f \funcT(\Lambda_f)$, $\Lambda_f{=}\sum_c \xcf$, and substitute it by corresponding weighted basepoint sums, the SOS$2$ constraint, and a convex combination.
First, we focus on one facility location and then add indexes to model all locations.
For location $f$, function $\funcT$ computes the TiS (\ref{eq:linearize1}). 
With its linearised version $\funcTa$ (\ref{eq:linearize2}) the convex part of the objective function,  $\Lambda_f \funcT(\Lambda_f)$,  becomes $\Lambda_f \sum_s \beta_s z_s$.
As $\Lambda_f$ depends on decision variable $\xcf$, multiplying $\xcf$ with $z_s$ turns the replacement term to be quadratic; only function $\funcT$ was linearised, not the whole objective function. 
However, having a linear and not quadratic objective function would reduce problem complexity and speeds up solving.
The quadratic term $\Lambda_f \sum_s \beta_s z_s$ needs to be replaced with an equivalent linear term.
This is achieved by ``moving'' the weight $\Lambda_f$ into function $\funcT$, which becomes $\funcT^w$~\eqref{eq:linearize3}.
Using $\funcTa^w$'s basepoints will transform the quadratic into the linear term $\sum_s \beta_s z_s$; the ordinate basepoints are now already weighted.
Since the other parts of the objective function were linear, the whole objective function is now linear.
\begin{IEEEeqnarray}{ll}
\funcT(x) = \frac{1}{\mu{-}x} = y, \textnormal{ with } x {=}\Lambda_f \label{eq:linearize1}\qquad \\
\funcTa : x=\sum_s z_s\alpha_s,\, \tilde y = \sum_s z_s\beta_s \label{eq:linearize2}\qquad \\
\funcT^w(x) = x\funcT(x) = \frac{x}{\mu{-}x} = y, \textnormal{ with } x {=}\Lambda_f \label{eq:linearize3}\qquad
\end{IEEEeqnarray}
To model all locations, index $f$ is added for each facility location forming the decision variables $z_{fs}$ and basepoint variables $\alpha_{fs}$, $\beta_{fs}$.
The linearised version $\optQPa(G,\, \lambda_c,\, \mu_f,\, p ,\, \alpha_s ,\, \beta_s)$ is hence:~%
\begin{IEEEeqnarray}{lllr}
\min_{x,y,z} \quad & \IEEEeqnarraymulticol{3}{l}{
    \frac{1}{\Lambda} \sum_{cf} \xcf \lcf + \frac{1}{\Lambda} \sum_{fs} \beta_{fs} z_{fs} 
}  \label{eq:qpa:obj} \\
\textnormal{s.t.} 
&    \sumf \xcf = \lambda_c, \qquad & \forall c \; & \textnormal{(demand)} \quad \label{eq:qpa:demand}\\
&    \sumc \xcf = \sums \alpha_{f s} \zfs, \quad & \forall f & \textnormal{(capacity)}\quad  \label{eq:qpa:capacity} \\
&    \sums \zfs = 1, \; \text{SOS}2(z_{f\cdot ,..}), \; & \forall f & \text{(pwl)}\quad  \label{eqa:pwl} \\
&    \sumc \xcf \leq y_f, & \forall f & \textnormal{(force flip)}\quad \label{eq:qpa:flip} \\
&    \sumf y_f = p & & \textnormal{(limit)}\quad  \label{eq:qpa:limit}
 \end{IEEEeqnarray}
The demand-weighted TiS is represented by term $\sums \beta_{fs} z_{fs}$ and the corresponding arrival rate at $f$ is $\sums \alpha_{fs} z_{fs} {=} \sum_f \Lambda_f {=} \sumc \xcf$; the new capacity constraint (\ref{eq:qpa:capacity}). 
This capacity constraint also implicitly  assures that the queuing system is in a steady state through the upper bound of the linearization interval $\alpha_{(m-1)} < \mu$; $\tau$ from the old constraint~(\ref{eq:qp:capacity}) becomes obsolete.

The search space of $\optQP$ consists of $|F|$ binary and $|F||C|$ real variables.
In addition $\optQPa$ has $m|F|$ real, restricted SOS variables.
If both problems were linear we could guess that solving the second problem $\optQPa$ takes longer than $\optQP$ because the search space is larger. 
However, linear problems are usually solved faster than non-linear problems.
Which problem is solved faster? The answers is not obvious, \eg, $\optQPa$ is linear but has a larger search space.
Their runtimes are experimentally evaluated in \Cref{sec:eval:lincon}.

The maximal linearisation error~\eqref{eq:opterror} of the objective function~\eqref{eq:qpa:obj} depends on the errors of the linearised parts, which is the sum of used resources, $y_f{=}1$, and their maximal linearisation errors $\epsilon_{\tilde{T}^w_\mu}$. 
\begin{IEEEeqnarray}{lllr}
\frac{1}{\Lambda} \sum_{f,\, y_f=1} \epsilon_{\tilde{T}^w_{\mu_f}} \; \leq \; 
\frac{p}{\Lambda} \max_{\tilde{T}^w} \left\lbrace \epsilon_{\tilde{T}^w } \right\rbrace \label{eq:opterror}
\end{IEEEeqnarray}
The linearisation accuracy drops if more resources are allowed to open ($p$). 
To maintain the same linearisation accuracy while doubling $p$ the linearisation error $\epsilon_{\tilde{T}^w}$ has to be halved.
This can be achieved by using more basepoints for the linearisation. 
Even if \Cref{fig:eval-pwl} indicates that less than twice the basepoints are necessary, increasing the number of basepoints and, hence, increasing the problem's search space increases the runtime.

\section{Evaluation} \label{sec:evaluation}

This section has four parts. 
First, it presents different TiS function linearisations to find a balance of two conflicting goals:
Small objective function approximation error and few basepoints ($m$) for fast computation.
Second, solutions of the convex and linear problem ($\optQP$ \vs $\optQPa$) are compared  for different real networks.
Third, the trade-off between the number of used locations and resulting response time is discussed.
Finally, application and network properties are presented for which considering the QD yields better response times than ignoring QD ($\optQPa$ \vs $\optP$).

\subsection{Weighted TiS Linearization}\label{sec:eval:lin}

This sections describes how we obtain the concrete basepoints for $\funcT^w(\lambda)$~\eqref{eq:linearize3} in the evaluation.
For this, we show a simplification with one set of basepoints adapted at runtime for different $\mu$ values.
Afterwards, we discuss the trade-off between a fast solving time and low approximation error.

Function $\funcT^w(\lambda)$ depends on $\mu$ and needs individual linearizations for different $\mu$; let $\alpha^\mu_s,\, \beta^\mu_s$ be their basepoints~\eqref{eq:wtis1}.
Alternatively, function $\textnormal{T}^w(\rho)$~\eqref{eq:wtis2} is independent of $\mu$ with corresponding basepoints $\alpha_s,\, \beta_s$.
Function $\funcT^w$ can be rewritten as $\textnormal{T}^w(\nicefrac \lambda \mu) {=} \funcT^w(\lambda)$~\eqref{eq:wtis3} and the corresponding basepoints can be rewritten similarly: $\forall s: \alpha^\mu_s {=} \mu \alpha_s , \, \beta^\mu_s {=} \beta_s$.
\bgroup
\small
\begin{IEEEeqnarray}{ll}
\funcT^w(\lambda) = \frac{\lambda}{\mu{-}\lambda}:  \quad  & 
    \sum_s \alpha^\mu_s z_s {=} \lambda,\, \sum_s \beta^\mu_s z_s {=} {\funcT^w}(\lambda) \qquad \label{eq:wtis1} \\
\textnormal{T}^w(\rho) = \frac {\rho} {1{-}\rho}: \quad &
   \sum_s \alpha_s z_s {=} \rho,\, \sum_s \beta_s z_s {=} {\textnormal{T}^w}(\rho) \qquad \label{eq:wtis2} \\
\textnormal{T}^w(\frac \lambda \mu)    =  \frac {\nicefrac \lambda \mu} {1{-\nicefrac \lambda \mu}} : \quad &
   \sum_s \alpha_s z_s {=} \frac \lambda \mu,\, \sum_s \beta_s z_s {=} {\textnormal{T}^w}(\frac \lambda \mu) \qquad \label{eq:wtis3} 
\end{IEEEeqnarray}
\egroup
As the ordinate basepoints $\beta_s$ remain unchanged, the basepoints' approximation error is also not affected.
With this handy transformation, we only need to precompute basepoints of $\textnormal{T}^w$ instead of basepoint sets of $\funcT^w$ for each different $\mu$ in the model, which speeds up the model setup process.

\begin{figure}[t]
\centering
\begin{inextended}
\includegraphics[width=9cm]{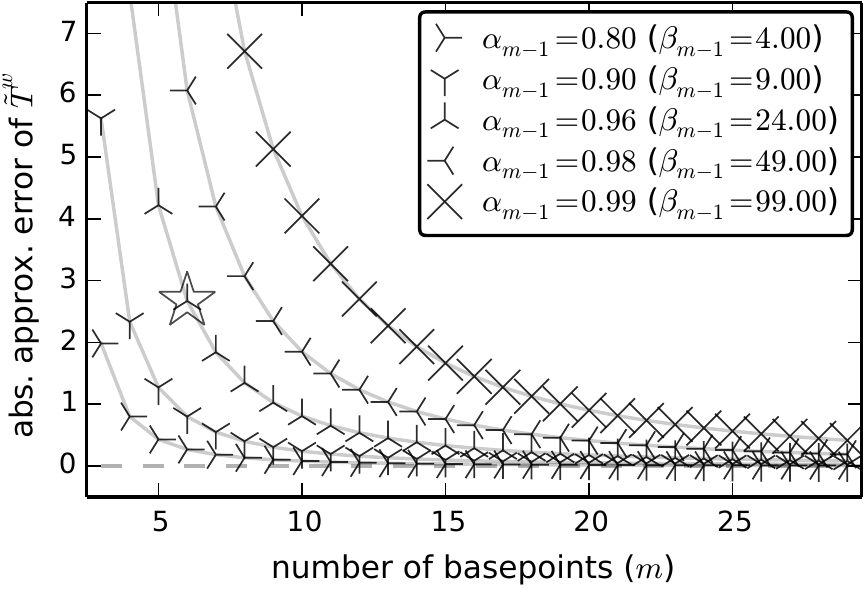}
\end{inextended}
\begin{notinextended}
\includegraphics[width=7cm]{./work/compare_pwlerror}
\end{notinextended}
\caption{
The error of linearising $\funcT^w$ is shown depending on the number of basepoints and different linearization intervals $[0, .., \alpha_{m{-}1}]$. 
}
\label{fig:eval-pwl}
\end{figure}
In the remaining section, we investigate the trade-off between a fast solving time and a low-error linearisation. 
For the first, we need to minimize the number of decision variables $z_{fs}$ or, equivalently, the number of basepoints used for the linearisation.
For the second, we investigate two control knobs (\Cref{sec:linearapprox}): many basepoints or small interval end $\alpha_{m-1}$.
\Cref{fig:eval-pwl} shows the error of $ \tilde{\textnormal{T}}^w$ depending on the number of basepoints $m$ for different $\alpha_{m-1}$ values.
We need a small error (down the vertical axis) with small $m$ (left on the horizontal axis) with large $\alpha_{m-1}$.
The latter also artificially limits the resource capacity and renders solving an input infeasible that could in fact be solved with larger $\alpha_{m-1}$.

For our evaluation, we set $\alpha_{m-1} {=} 0.96$ and $m{=}6$ with error $\epsilon_{\funcT^w} {=} 2.67$ as a good compromise between the number of decision variables, approximation error, and artificial capacity limit.

\subsection{Comparison: Convex vs. Linear} \label{sec:eval:lincon}

\begin{figure*}[tb]
\centering
\includegraphics[scale=0.9]{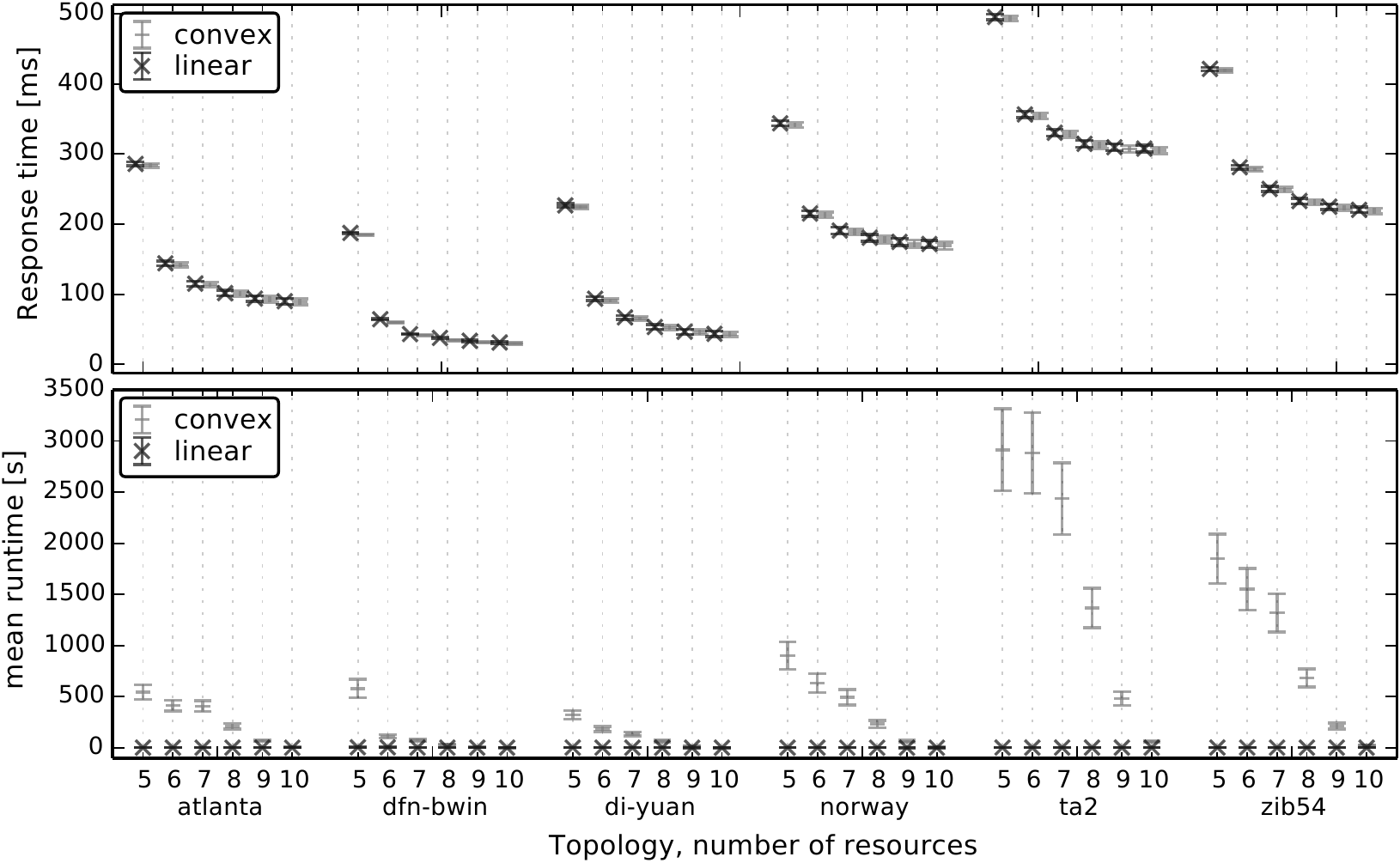}
\caption{
Solution quality (a) and solving time (b) for convex $\optQP$ and its linearizations $\optQPa$.
Estimated mean confidence intervals at $95\%$~confidence level.
}
\label{fig:deviation}
\end{figure*}

We choose the following structurally different topologies from SndLib~\cite{Orlowski2007}:
\texttt{ta2}, \texttt{zib54} with many nodes (around $50$); \texttt{yuan}, \texttt{bwin} with few nodes (around 10); \texttt{atlanta}, \texttt{norway} for dense networks (node:edge ratio 1:2).
All topologies are connected.
We approximate the latency between nodes by their geographical distance~\cite{Kaune2009}.
We assume that data centres are built at well connected nodes/routers, so we selected the $10$ nodes with the highest degree\footnotemark\ to be data centres.
\footnotetext{For same degrees the node id is the tie breaker.}
We set a relatively low service rate of $\mu{=}\SI{100}{\req\per\s}$ to reflect our computation-intensive example applications~\cite{Wan2010, Barker2010,Lee2012}. 
The service rates were the same for all data centres.
User requests arrive at all nodes and the arrival rate $\lambda_c$ for users at site $c$  is randomly generated: 
Each value is uniformly drawn from an $[0,1]$ interval and, afterwards, all values are normalized to $\sum_c\lambda_c{=}\SI{470}{\req\per\s}$. 
This value, together with the service rate $\mu{=}\SI{100}{\req\per\s}$, ensures feasibility for $5$ or more facilities.
We randomly generated $50$ different realizations for all arrival rates.
For each of these $50$ sets of arrival rates, we considered $p {\in} [5,\, 10 ]$ facilities, resulting in $300$ configurations per topology.
Each configuration was solved using either $\optQP$ or $\optQPa$.

\Cref{fig:deviation} compares the solution quality (a) and solving time (b). 
The horizontal axis lists the groups of different topologies and the used number of resources for each group.
The vertical axis in (a) shows the $95\%$~confidence intervals of the average response times as the quality of solutions obtained by solving $\optQP$ (cross) and $\optQPa$ (line) for each of the $50$~realizations for each group.
Similarly, the vertical axis in (b) shows the $95\%$~confidence intervals of the solving time for each group.
\emph{The $\optQPa$'s solutions have a very similar quality to $\optQP$'s solutions.}

Looking at (b), obtaining $\optQP$'s solutions took longer than obtaining the $\optQPa$'s solutions.
However, those values have to be interpreted with care. 
Our implementation of $\optQP$ has to process all possible combinations ($F'$), whereas $\optQPa$ benefits from the MIP solver's branch-and-cut algorithm to reduce the search space.
To compare this structurally different problems, we restrict the number of candidate facilities to $10$ and the number of possible combinations; the major cause of the higher solving time of $\optQP$.
Nevertheless, \emph{the absolute solving times of $\optQPa$ are very short} for all groups.

In conclusion, \emph{the linearised problem~$\optQPa$ is an adequate substitution for our original problem~$\optQP$: fast and accurate.}

\subsection{Response Time Reduction }\label{sec:eval:resptime}

\Cref{fig:deviation} also shows how response time improves when adding a resource.
We could verify two effects decreasing the response times: 
First, using more locations allows better load balancing, which reduces the queuing delays. 
These reductions are larger for highly utilized locations than for less utilized ones. 
Second, more locations allow nearer locations, reducing the round trip times.
In conclusion, the average response time of $\optQP(...,\,p)$ decreases monotonically in number of resources.
Then, service providers earning more money by connecting users with lower response times face a diminishing return. 
At one breaking point~$p^*$ the cost for adding a resource will exceed the additionally earned money.
This point depends on the topology, service times, and service monetization.
By using $\optQP$ with different $p$ values, the service provider can determine $p^*$ in advance to avoid profit loss.

\begin{inextended}
\begin{figure*}[tb]
\centering
\includegraphics[scale=0.9]{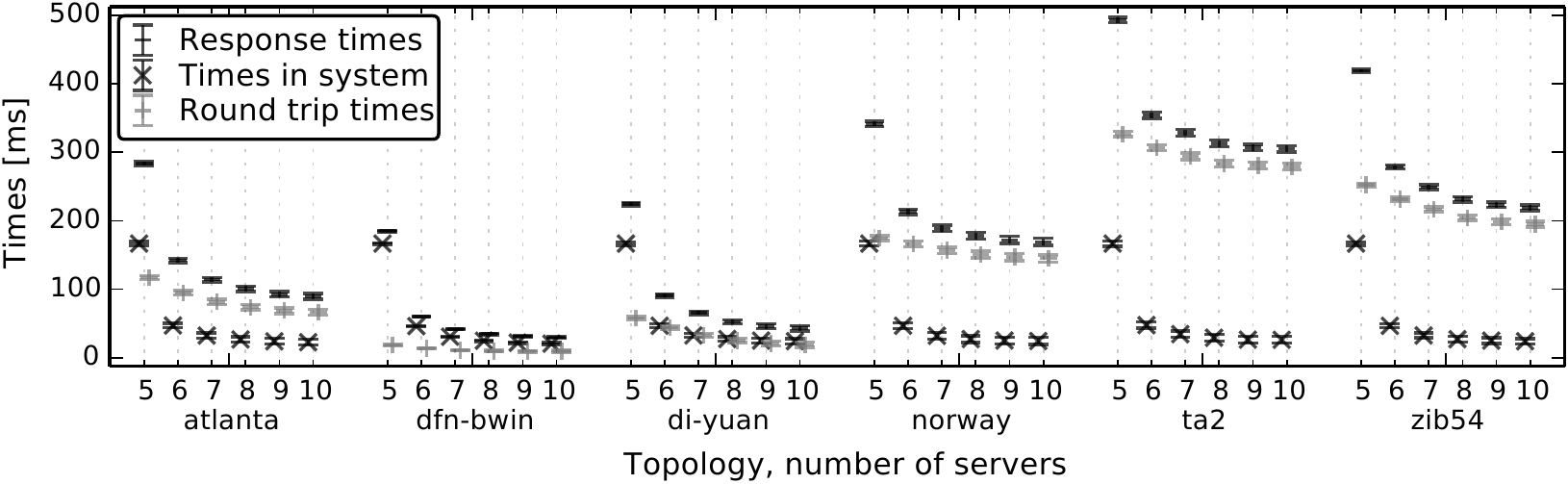}
\caption{
Round trip times, times in system, and round trip times for different topologies and numbers of used servers.
Estimated mean confidence intervals at $95\%$~confidence level.
}
\label{fig:comparert}
\end{figure*}
For a closer look, \Cref{fig:comparert} shows not only the average response time but also the time in system and round trip times grouped along the horizontal axis the same way like in~\Cref{fig:deviation}. 
We can trace the two effects of response time reductions: 
First, load balancing across more resources reduces the queuing delay and with it the time in queuing system.
Second, the round trip time is reduced as nearer resources are used.
\end{inextended}

\subsection{Considering Queueing Delays} \label{sec:eval:qtornot}

\begin{inextended}
\begin{figure*}[tbp]
\centering
\includegraphics[scale=1.2]{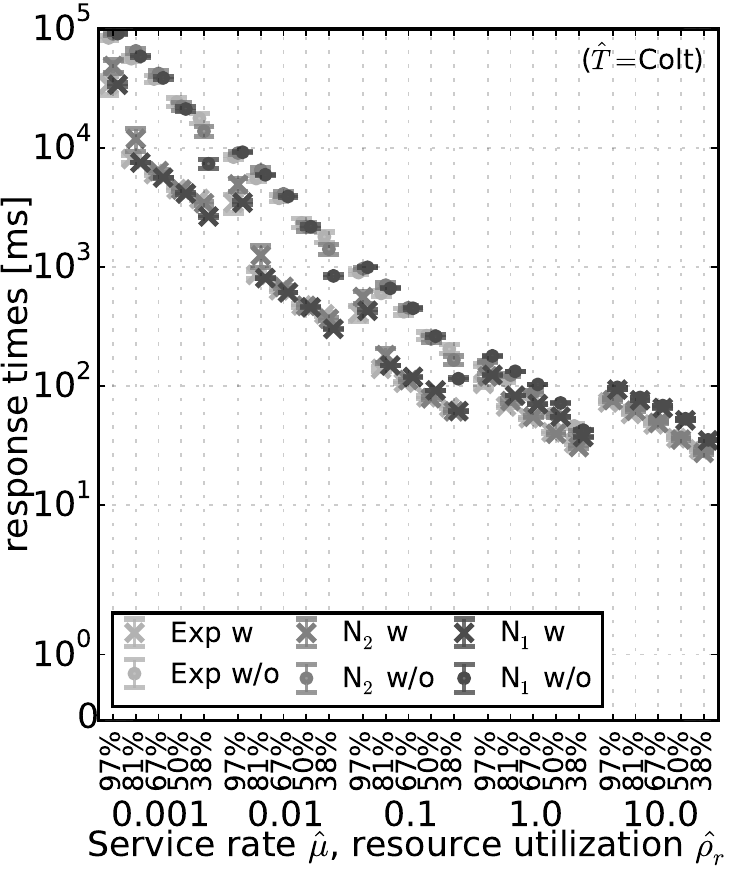}
\includegraphics[scale=1.2]{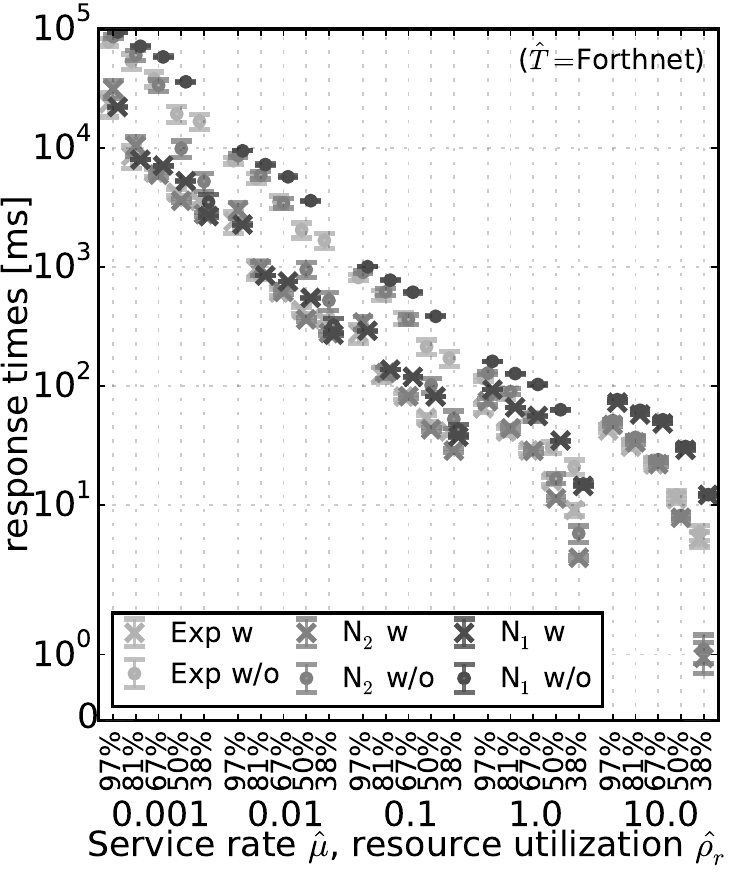}
\caption{Average response times as a function of factor combination ($ \hat{\mu} $, $ \hat{\rho_s} $) for different demand distributions ($\hat{D}$) with $95\%$~confidence intervals. 
Left (and right) comparison for topology $\hat{T} {=} $\texttt{Colt} (and \texttt{Forthnet}, respectively). \label{fig:rt:cicompare} }
\end{figure*}
\end{inextended}

\begin{notinextended}
\begin{figure*}[tbp]
\centering
    \begin{minipage}{.48\textwidth}
    \centering
    \includegraphics[scale=1.]{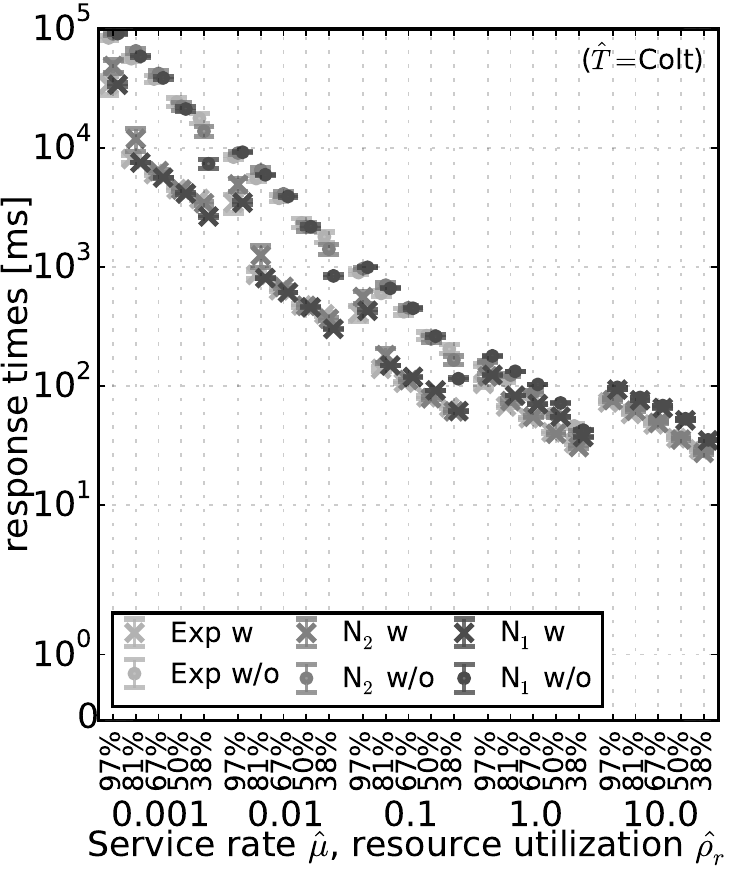}
    \end{minipage}\hfill
    \begin{minipage}{.5\textwidth}
    \centering
    \includegraphics[scale=1.]{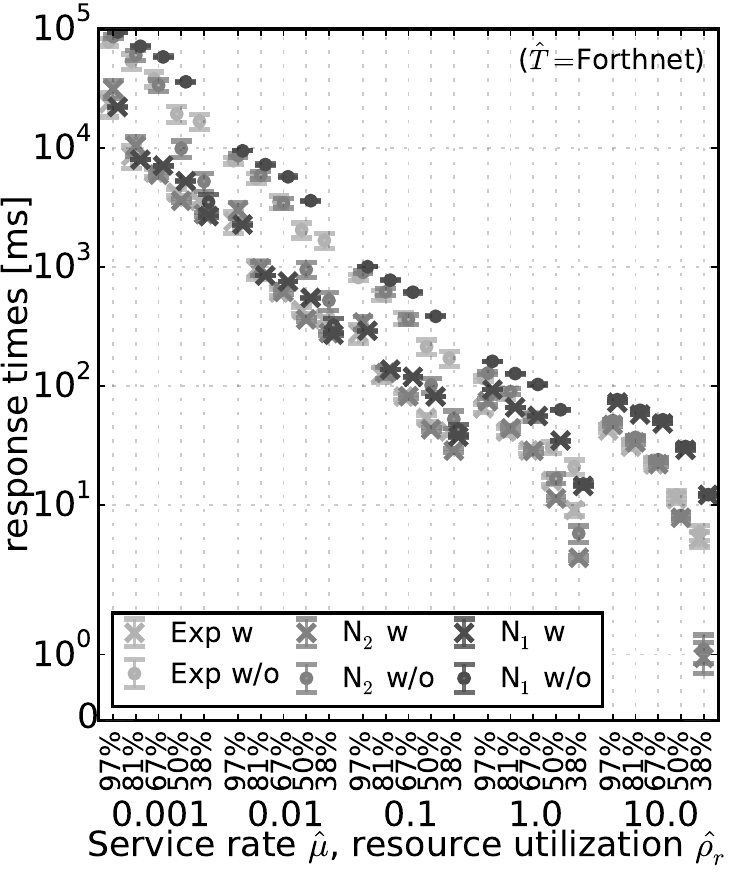}
    \end{minipage}    
\caption{Average response times as a function of factor combination ($ \hat{\mu} $, $ \hat{\rho_s} $) for different demand distributions ($\hat{D}$) with $95\%$~confidence intervals. 
Left (and right) comparison for topology $\hat{T} {=} $\texttt{Colt} (and \texttt{Forthnet}, respectively). \label{fig:rt:cicompare} }
\end{figure*}
\end{notinextended}

This paper presents a refinement of the assignment problem ($\optP$ in~\eqref{eq:p:obj}) by additionally considering queuing delays. 
This refinement increases problem's complexity, accuracy, and solving time.
Only a significantly lower response time would make these drawbacks worthwhile; for instances, small queuing delays compared to  large round trip times will render the refinement unnoticeable.
This section investigates multiple scenario factors influencing queuing delays and judges the refinement gain by comparing 
solutions' response times obtained by either ignoring or considering queuing delays.

For this, we perform a second experiment with a slightly different configuration as in \Cref{sec:eval:lincon}.
The first experiment showed that $\optQPa$ is a very accurate and fast substitution for $\optQP$.
Focussing on $\optQPa$ allows us to evaluate more scenario variations in a reasonable time than would be possible with $\optQP$.

\subsubsection{Configuration}
We vary configurations by four factors\footnotemark\ $\hat{\mu}$, $\hat{D}$, $\hat{\rho}_s$, $\hat{G}$.%
\footnotetext{The $\hat\ $ indicates the factors under investigation.}
The service rate $\hat{\mu}$ reflects how computation-intensive the application is. 
A small service rate means a high processing time. 
For the same arrival rate, queuing delays are higher for smaller service rates.
For clarity, we assume homogeneous\footnotemark\ resources, same service times at each location, $\hat \mu {=} \mu_f,\,\forall f$. %
\footnotetext{
We discard an additional minimization potential of heterogeneous resources to simplify the comparison:
When assigning demand, a farer but faster resource enables trading off a larger RTT for a smaller QD+PT. By doing so, the response time is further reduced.}%
Different levels of $\hat{\mu}$ represents different application types ranging from fast web servers with a short processing time up to computation-intensive applications, $\hat \mu {=} \SIlist[list-final-separator={; }]{10000; 1000; 100; 10; 1}{\req\per\s}$.

Unlike the service rates, the arrival rates~$\lambda_c$ are not homogeneous but randomly distributed.
This enables us to investigate different patterns for spatially distributed load, \eg, fluctuations or local hot spots.
Let $\hat{\lambda}$ be the targeted mean arrival rate.
For this, we choose three different random distributions~$\hat D$ for our second factor.
First, a similar load across all nodes with small fluctuations is represented by a ``narrow'' normal distribution: $\textnormal{N}(\textnormal{mean}{=}\hat\lambda,\,\textnormal{std.dev.}{=}\nicefrac {\hat\lambda}{20} ) {=} \textnormal{N}_1$.
Second, a largely fluctuating load around an average load per node is represented by a ``wide'' normal distribution:
$\textnormal{N}(\hat\lambda,\,\hat\lambda){=}\textnormal{N}_2$.
Third, heavy variations causing local hot spots are represented by an exponential distribution:
$\textnormal{Exp}(\hat{\lambda})$.
For each node the arrival rate~$\lambda_c$ is drawn from $\hat D$, where negative values are capped to zero, $\lambda_c{=}\max\{0,X\},\,X{\sim}\hat D {\in} \{\textnormal{N}_1, \textnormal{N}_2, \textnormal{Exp} \}$. 
We investigate $50$ different such realizations for each topology.

The third factor $\hat{\rho}_r$ reflects the average resource utilization.
Highly utilized resources have high QDs, and solutions are very similar whether or not the queuing delay is considered.
To enforce the $\hat{\rho}_r$ levels, we limit the number of utilised resources $p$ defined later.

The last factor, with different topologies $\hat G$ we represents structural differences like the diameter or ratio between the round trip time and queuing delays, for instance, sparse graphs have a larger diameter and higher round trip times than dense graphs. 
We selected $14$ out of $524$ topologies from different sources: 
sndlib\footnote{\label{fn:lat} Round trip times were approximated by geographical distances~\cite{Kaune2009}.}~\cite{Orlowski2007}, 
topology zoo\cref{fn:lat}~\cite{Knight2011}, %
kingtrace\footnotemark~\cite{Gummadi2002}.
The selection considers three categories for the number of nodes, edge, and diameter to eliminate roughly similar topologies. 
\footnotetext{A sparse matrix specifies point to point latencies. Some were only available in one direction. We assume the same latency for the opposite direction; otherwise those nodes had to be discarded.}

In each topology, the \num{100} best connected nodes\footnotemark\ were marked as candidate resource locations $F$, $|F| {=} \min\left\lbrace |N|, 100 \right\rbrace$.
\footnotetext{Highest degree first; node ID as the tie breaker.}
Let $p_{\textnormal{min}} \leq |F|$ be the number of resources at least necessary to handle all demand.
Having more demand using more resources means less freedom for location choices: for instance,  if $p_{\textnormal{min}} {=} |F|$ all resources are fully utilized and only this decision is possible. 
We set $p_{\textnormal{min}} {=} \lfloor 0.3 |F| \rfloor$ to allow enough freedom.
Then, $\hat D$'s target arrival rate~$\hat{\lambda}$ is defined accordingly, $\hat{\lambda} {=} \hat \mu \frac {p_{\textnormal{min}} }{|N|}$. 
Actually using  $p{=}p_{\textnormal{min}}$ resources results in a very high resource utilization~$p_r$; allowing more resources reduces the utilization.
For our third factor, the resource utilization $\hat \rho_r{=} \numlist{0.97;0.81;0.67; 0.5; 0.375}$, we set corresponding $p$ values achieving (roughly) the targeted server utilization, $\hat p{=} \lfloor \nicefrac{0.3}{\hat \rho_r}|F| \rfloor {=} \lfloor a|F|\rfloor$, $a {=} \numlist{0.31;0.37;0.45; 0.6; 0.8}$.

\subsubsection{Results} 

Summarising, each of the \num{1050} combinations of the four factors ($\hat \mu$, $\hat D$, $\hat \rho_s$, $\hat T$) was randomised by \num{50} demand realisations, resulting in \num{52500} different configurations for which problem~$\optP$ (without considering queuing delays) and problem~$\optQPa$ (with considering queuing delays) are solved.
The quality metric is the average response time computed by $\optQP$'s exact objective function \eqref{eq:qp:obj}.
The difference between response time obtained with $\optQPa$ and with $\optP$ measures the response time improvement when considering queuing delays. 
A larger difference means $\optQPa$'s assignments are superior to $\optP$'s assignments. 
As $\optP$ is a simplification of $\optQP$, $\optP$'s response times (RTs) cannot be smaller than $\optQP$'s RTs. 
Only using the linearisation~$\optQPa$ could potentially result in a higher RT, but never occur\footnotemark{} in our evaluation.
\footnotetext{Some pairs of $\optQPa$ and $\optP$ solutions are treated as equivalent good when their response time differences were below $10^{-6}$ -- the solver threshold at which the solver stopped improving the solution in our evaluation.}

\Cref{fig:rt:cicompare} shows response times as a function of two factors, service rate~$\hat \mu$ and resource utilisation~$\hat \rho_s$ for two selected topologies $\hat T{=}$\texttt{Colt}, \texttt{Forthnet}.
Each single data point corresponds to the average response times with $95\%$~confidence intervals of $50$~realisations for one factor combination $(\hat{\mu}, \hat{\rho_s}, \hat{T})$.
At a first glance, the response times of the two problems can be compared column-wise: 
The circle data points represent problem~$\optP$'s solutions. 
Most of them are above the cross data points representing the $\optQPa$'s solutions, no circle appears above a cross:
The difference between a column-wise cross circle pair visualizes the response time improvements when considering queuing delays; the $y$-axis is in log-scale.

At a second look, \Cref{fig:rt:cicompare} shows another pattern along increasing service rates ($\hat{\mu}$) from \SI{1}{\req\per\ms} to \SI{10000}{\req\per\ms} (left to right): Significant response time improvements for low service rates and nearly no response time improvements for high service rates.
\emph{When queuing delay (and processing times) become a significant part of the response time, it is necessary to consider queuing delays when deciding the assignments} as otherwise unnecessary higher response times would be the result.

At a third look, \Cref{fig:rt:cicompare} shows larger response time differences for higher utilized ($\hat{\rho_s}$) configurations; in these cases, the queuing delay becomes a dominant part of the response time. \emph{Considering queuing delays is necessary for utilised topologies ($\rho_s{>}0.5$).}

At a fourth look, \Cref{fig:rt:cicompare} shows how the demand distributions represented by different grey levels influence the response time: While the \texttt{Colt} topology show few changes in response times for different demand distributions, the \texttt{Forthnet} topology data points are more spread for different demand distributions. 
In tendency, the more challenging demand distributions (Exp and N$_2$) show larger response time improvements than the other distribution (N$_1$).
Considering queuing delays avoids assigning local demand hotspots to nearby but highly utilized resources; instead more and lesser utilized resources are used effectively reducing the average response time.

In summary, \emph{the evaluation verifies our claim, that considering queuing delays is necessary as response times can significantly reduced.} This applies only when queuing delays and round trip times are of the same magnitude; \eg configuration with service rate~$\hat{\mu}{=}\SI{10000}{\req\per\ms}$ (short queuing delays) have only marginal response time improvements when considering queuing delays.

\section{Conclusion}
We extend previous work by optimally solving the assignment problem $\optQP$, a Facility Location Problem with integrated queuing systems. 
We proposed problem~$\optQPa$ as a linearisation of $\optQP$ with accurate solutions obtained fast.
In our scenario of adapting the resource allocation at geographically distributed sites, such a swift reaction is important. 
It allows to swiftly react immediately to an ever changing environment including demand fluctuations or network congestion.

We showed that adding more and more resources will at one point reduce the user expected response time only marginally. With our work, the application provider can determine this point in advance and can allocate resources accordingly.

We performed a large-scale experiment and traced down network and application properties where integrating the queuing system into the FLP improves solutions. This could guide other researchers or application providers whether the complex problem~$\optQP$ is necessary to apply or the simpler problem~$\optP$ is sufficient enough for their scenario.

The simple M/M/1-queue model can be replaced with more sophisticated queuing models as long as (i)~the inter-arrival times are described by a Poisson process and (ii)~the queuing delay function is convex. 
For models with a different inter-arrival time process, splitting and joining becomes much more complicated.
The linearised problem even supports non-convex queuing delay functions, but for such models the presented algorithm for obtaining basepoints is no more applicable.

Finally, the shown modelling techniques can be used beyond our use case.
The assignment problem $\optQP$ is part of the broader family of \emph{FLPs with convex cost functions}.
We think for most, maybe all of them, a similar good and fast problem linearisation can be formulated by reusing our problem linearization formulation and by determining the PWL basepoints with the presented algorithm.

\bibliographystyle{abbrv}
\bibliography{mendexport2}

\begin{thebibliography}{10}

\bibitem{aws}
{Amazon Web Service}.

\bibitem{Aboolian2009}
R.~Aboolian, O.~Berman, and Z.~Drezner.
\newblock {The multiple server center location problem}.
\newblock {\em Annals of Operations Research}, 167(1):337--352, mar 2009.

\bibitem{Agarwal2010}
S.~Agarwal, J.~Dunagan, and N.~Jain.
\newblock {Volley: Automated data placement for geo-distributed cloud
  services}.
\newblock In {\em Proceedings of the 7th conference on Networked systems design
  and implementation (NSDI '10)}. USENIX, 2010.

\bibitem{Alicherry2012}
M.~Alicherry and T.~Lakshman.
\newblock {Network aware resource allocation in distributed clouds}.
\newblock In {\em 2012 Proceedings IEEE INFOCOM}, pages 963--971. IEEE, mar
  2012.

\bibitem{cvxopt}
M.~S. Andersen and L.~Vandenberghe.
\newblock {CVXOPT - a free software package for convex optimisation}, 2013.

\bibitem{Bagaa2014}
M.~Bagaa, T.~Taleb, and A.~Ksentini.
\newblock {Service-aware network function placement for efficient traffic
  handling in carrier cloud}.
\newblock In {\em Proceedings of the Wireless Communications and Networking
  Conference (WCNC)}, pages 2402--2407. IEEE, apr 2014.

\bibitem{Barker2010}
S.~K. Barker and P.~Shenoy.
\newblock {Empirical evaluation of latency-sensitive application performance in
  the cloud}.
\newblock In {\em Proceedings of the 1st annual conference on Multimedia
  systems (MMSys)}, New York, 2010. ACM Press.

\bibitem{Bauer2011}
M.~Bauer, S.~Braun, and P.~P. Domschitz.
\newblock {Media Processing in the Future Internet}.
\newblock In {\em Proceedings of the 11th W{\"{u}}rzburg Workshop on IP:
  Visions of Future Generation Networks}, pages 113--115, W{\"{u}}rzburg, 2011.

\bibitem{Beale1970}
E.~Beale, M.~Lansdowne, and J.~A. Tomlin.
\newblock {Special facilities in a general mathematical programming system for
  non-convex problems using ordered sets of variables}.
\newblock {\em Operational Research}, 69(447-454):99, 1970.

\bibitem{Berman2006}
O.~Berman and Z.~Drezner.
\newblock {The multiple server location problem}.
\newblock {\em Journal of the Operational Research Society}, 58(1):91--99,
  2007.

\bibitem{Bolch2005}
G.~Bolch, S.~Greiner, H.~{De Meer}, and K.~S. Trivedi.
\newblock {\em {Queueing networks and Markov chains: Modeling and Performance
  Evaluation with Computer Science Applications}}.
\newblock Wiley-Interscience, 2005.

\bibitem{Boyd2004}
S.~Boyd and L.~Vandenberghe.
\newblock {\em {Convex optimization}}, volume~25.
\newblock Cambridge University Press, jun 2004.

\bibitem{Cai2013}
D.~Cai and S.~Natarajan.
\newblock {The Evolution of the Carrier Cloud Networking}.
\newblock In {\em Seventh International Symposium on Service-Oriented System
  Engineering}, pages 286--291. IEEE, mar 2013.

\bibitem{Cardellini2000}
V.~Cardellini, M.~Colajanni, and P.~Yu.
\newblock {Geographic load balancing for scalable distributed Web systems}.
\newblock In {\em Proceedings of the 8th International Symposium on Modeling,
  Analysis and Simulation of Computer and Telecommunication Systems}, pages
  20--27. IEEE Comput. Soc, 2000.

\bibitem{Church2008}
K.~Church, A.~Greenberg, and J.~Hamilton.
\newblock {On Delivering Embarrassingly Distributed Cloud Services}.
\newblock In {\em HotNets}, 2008.

\bibitem{Claypool2006}
M.~Claypool and K.~Claypool.
\newblock {Latency and player actions in online games}.
\newblock {\em Communications of the ACM - Entertainment networking},
  49(11):40--45, 2006.

\bibitem{Csaszar2013a}
A.~Csaszar, W.~John, M.~Kind, C.~Meirosu, G.~Pongracz, D.~Staessens, A.~Takacs,
  and F.~J. Westphal.
\newblock {Unifying cloud and carrier network: EU FP7 Project UNIFY}.
\newblock In {\em 6th International Conference on Utility and Cloud Computing},
  pages 452--457. IEEE/ACM, 2013.

\bibitem{Cucinotta2013}
T.~Cucinotta, K.~Oberle, M.~Stein, P.~Domschitz, and S.~Mullender.
\newblock {Run-time Support for Real-Time Multimedia in the Cloud}.
\newblock In {\em 2nd International Workshop on Real-Time and Distributed
  Computing in Emerging Applications (REACTION 2013)}, 2013.

\bibitem{Drezner2011}
T.~Drezner and Z.~Drezner.
\newblock {The gravity multiple server location problem}.
\newblock {\em Computers {\&} Operations Research}, 38(3):694--701, mar 2011.

\bibitem{Drezner2004}
Z.~Drezner and H.~W. Hamacher.
\newblock {\em {Facility location: applications and theory}}.
\newblock Springer, 2004.

\bibitem{Endo2011}
P.~T. Endo, A.~{de Almeida Palhares}, N.~Pereira, G.~Goncalves, D.~Sadok,
  J.~Kelner, B.~Melander, and J.-E. Mangs.
\newblock {Resource allocation for distributed cloud: concepts and research
  challenges}.
\newblock {\em IEEE Network}, 25(4):42--46, 2011.

\bibitem{Fischer2013}
A.~Fischer, J.~F. Botero, M.~T. Beck, H.~de~Meer, and X.~Hesselbach.
\newblock {Virtual Network Embedding: A Survey}.
\newblock {\em Communications Surveys {\&} Tutorials, IEEE}, (99):1--19, 2013.

\bibitem{Freeman2006}
M.~J. Freeman, K.~Lakshminarayanan, and {David Mazieres}.
\newblock {OASIS: Anycast for Any Service}.
\newblock In {\em Networked Systems Design and Implementation (NSDI)}, 2006.

\bibitem{MINLP2012}
B.~Geissler, A.~Martin, A.~Morsi, and L.~Schewe.
\newblock {Using Piecewise Linear Functions for Solving MINLPs}.
\newblock In J.~Lee and S.~Leyffer, editors, {\em Mixed Integer Nonlinear
  Programming}, volume 154 of {\em The IMA Volumes in Mathematics and its
  Applications}, pages 287--314. Springer New York, 2013.

\bibitem{Goudarzi2013}
H.~Goudarzi and M.~Pedram.
\newblock {Geographical Load Balancing for Online Service Applications in
  Distributed Datacenters}.
\newblock In {\em 6th International Conference on Cloud Computing}, pages
  351--358. IEEE, 2013.

\bibitem{Gummadi2002}
K.~P. Gummadi, S.~Saroiu, and S.~D. Gribble.
\newblock {King: Estimating Latency between Arbitrary Internet End Hosts}.
\newblock In {\em Proceedings of the 2nd Workshop on Internet measurment (IMW
  '02)}, pages 5--18. ACM SigComm, 2002.

\bibitem{Hakimi1979}
O.~Hakimi and S.~L. Kariv.
\newblock {An Algorithmic Approach to Network Location Problems. II: The
  p-Medians}.
\newblock {\em SIAM Journal on Applied Mathematics}, 37(3):539--560, feb 1979.

\bibitem{Hao2009}
F.~Hao, T.~V. Lakshman, S.~Mukherjee, and H.~Song.
\newblock {Enhancing dynamic cloud-based services using network
  virtualization}.
\newblock {\em Proceedings of the 1st ACM workshop on Virtualized
  infrastructure systems and architectures (VISA)}, 40(1):37, 2009.

\bibitem{Imamoto2008}
A.~Imamoto and B.~Tang.
\newblock {Optimal Piecewise Linear Approximation of Convex Functions}.
\newblock In {\em World Congress on Engineering}, pages 22----25, 2008.

\bibitem{Ishii2011}
A.~Ishii and T.~Suzumura.
\newblock {Elastic Stream Computing with Clouds}.
\newblock In {\em Proceedings of the 4th International Conference on Cloud
  Computing}, pages 195--202. IEEE, jul 2011.

\bibitem{Algorithmic1979}
O.~Kariv and S.~L. Hakimi.
\newblock {An algorithmic approach to network location problems. The
  p-medians}.
\newblock {\em SIAM Journal on Applied Mathematics}, 37(3):539--560, 1979.

\bibitem{Kaune2009}
S.~Kaune, K.~Pussep, C.~Leng, A.~Kovacevic, G.~Tyson, and R.~Steinmetz.
\newblock {Modelling the internet delay space based on geographical locations}.
\newblock In {\em Proceedings of the 17th Euromicro International Conference on
  Parallel Distributed and Network-based Processing}, pages 301--310. IEEE,
  2009.

\bibitem{Keller2014}
M.~Keller and H.~Karl.
\newblock {Response Time-Optimized Distributed Cloud Resource Allocation}.
\newblock In {\em Workshop on Distributed Cloud Computing (DCC)}. ACM, 2014.

\bibitem{Keller2013b}
M.~Keller, M.~Peuster, C.~Robbert, and H.~Karl.
\newblock {A Topology-aware Adaptive Deployment Framework for Elastic
  Applications}.
\newblock In {\em 17th International Conference on Intelligence in Next
  Generation Networks (ICIN)}, pages 61--69. IEEE, oct 2013.

\bibitem{Knight2011}
S.~Knight, H.~X. Nguyen, N.~Falkner, R.~Bowden, and M.~Roughan.
\newblock {The Internet Topology Zoo}.
\newblock {\em Journal on Selected Areas in Communications}, 29(9):1765--1775,
  oct 2011.

\bibitem{Kwon2014}
M.~Kwon.
\newblock {A Tutorial on Network Latency and its Measurements}.
\newblock {\em IGI Global}, 2014.

\bibitem{Lee2012}
Y.-T. Lee, K.-T. Chen, H.-I. Su, and C.-L. Lei.
\newblock {Are all games equally cloud-gaming-friendly? an electromyographic
  approach}.
\newblock In {\em Proceedings of the 11th Annual Workshop on Network and
  Systems Support for Games (NetGames)}, pages 4--9. ACM/IEEE, 2012.

\bibitem{Lin2012}
M.~Lin, Z.~Liu, A.~Wierman, and L.~L.~H. Andrew.
\newblock {Online algorithms for geographical load balancing}.
\newblock In {\em International Green Computing Conference (IGCC)}, pages
  1--10, 2012.

\bibitem{Liu2015}
Z.~Liu, M.~Lin, A.~Wierman, S.~Low, and L.~Andrew.
\newblock {Greening Geographical Load Balancing}.
\newblock {\em Transactions on Networking (TON)}, 23(2):657--671, 2015.

\bibitem{Marianov1996}
V.~Marianov and D.~Serra.
\newblock {Probabilistic maximal covering location-allocation models with
  constrained waiting time or queue length for congested systems}.
\newblock {\em Journal of Regional Science}, 38(3):401--424, 1996.

\bibitem{Marianov2002}
V.~Marianov and D.~Serra.
\newblock {Location – Allocation of Multiple-Server Service Centers}.
\newblock {\em Annals of Operations Research}, 111(1--4):35----50, 2002.

\bibitem{Moghadas2011}
M.~Moghadas and T.~Kakhki.
\newblock {Maximal covering location-allocation problem with M/M/k queuing
  system and side constraints}.
\newblock {\em Iranian Journal of Operations Research}, 2(2):1--16, 2011.

\bibitem{Oberle2013}
K.~Oberle, D.~Cherubini, and T.~Cucinotta.
\newblock {End-to-end service quality for cloud applications}.
\newblock In {\em Economics of Grids, Clouds, Systems, and Services}, volume
  8193 LNCS, pages 228--243. Springer, 2013.

\bibitem{Orlowski2007}
S.~Orlowski, M.~Pioro, A.~Tomaszewski, and R.~Wessaly.
\newblock {SNDlib Survivable Network Design Library}.
\newblock Technical Report June, Konrad-Zuse-Zentrum f{\"{u}}r
  Informationstechnik Berlin, Spa, Belgium, 2007.

\bibitem{Padberg2000}
M.~Padberg.
\newblock {Approximating separable nonlinear functions via mixed zero-one
  programs}.
\newblock {\em Operations Research Letters}, 27(1):1--5, aug 2000.

\bibitem{Pandey2010}
S.~Pandey, A.~Barker, K.~K. Gupta, and R.~Buyya.
\newblock {Minimizing Execution Costs when Using Globally Distributed Cloud
  Services}.
\newblock In {\em Proceedings of the 24th International Conference on Advanced
  Information Networking and Applications}, pages 222--229. IEEE, 2010.

\bibitem{Pasandideh2010}
S.~H.~R. Pasandideh and S.~T.~A. Niaki.
\newblock {Genetic application in a facility location problem with random
  demand within queuing framework}.
\newblock {\em Journal of Intelligent Manufacturing}, 23(3):651--659, may 2010.

\bibitem{Scharf2012}
M.~Scharf, T.~Voith, W.~Roome, B.~Gaglianello, M.~Steiner, V.~Hilt, and V.~K.
  Gurbani.
\newblock {Monitoring and abstraction for networked clouds}.
\newblock In {\em Proceedings of the 16th International Conference on
  Intelligence in Next Generation Networks (ICIN)}, pages 80--85. IEEE, oct
  2012.

\bibitem{Shah2015}
B.~Shah and B.~H. Trivedi.
\newblock {Improving Performance of Mobile Agent Based Intrusion Detection
  System}.
\newblock {\em Fifth International Conference on Advanced Computing {\&}
  Communication Technologies}, pages 425--430, 2015.

\bibitem{Stolyar2013a}
A.~L. Stolyar and B.~Labs.
\newblock {Shadow-Routing Based Dynamic Algorithms for Virtual Machine
  Placement in a Network Cloud}.
\newblock In {\em Proceedings of the 32nd International Conference on Computer
  Communications (INFOCOM '13)}, pages 644--652. IEEE, 2013.

\bibitem{Taleb2014}
T.~Taleb.
\newblock {Toward carrier cloud: Potential, challenges, and solutions}.
\newblock {\em Wireless Communications, IEEE}, 21(3):80----91, jun 2014.

\bibitem{Taleb2013}
T.~Taleb and A.~Ksentini.
\newblock {Follow Me cloud: Interworking federated clouds and distributed
  mobile networks}.
\newblock {\em IEEE Network}, 27(5):12--19, 2013.

\bibitem{Vidyarthi2009}
N.~Vidyarthi, S.~Elhedhli, and E.~Jewkes.
\newblock {Response time reduction in make-to-order and assemble-to-order
  supply chain design}.
\newblock {\em IIE Transactions}, 41(5):448--466, mar 2009.

\bibitem{Wan2010}
Z.~Wan.
\newblock {Cloud Computing infrastructure for latency sensitive applications}.
\newblock In {\em Proceedings of the 12th International Conference on
  Communication Technology}, pages 1399----1402. IEEE, nov 2010.

\bibitem{Wang2002}
Q.~Wang, R.~Batta, and C.~M. Rump.
\newblock {Algorithms for a Facility Location Problem with Stochastic Customer
  Demand and Immobile Servers}.
\newblock {\em Annals of Operations Research}, 111(1-4):17--34, mar 2002.

\bibitem{Wendell2010}
P.~Wendell, J.~W. Jiang, M.~J. Freedman, and J.~Rexford.
\newblock {DONAR : Decentralized Server Selection for Cloud Services}.
\newblock In {\em SigComm}, volume~40, pages 231--242, 2010.

\bibitem{Wong2006}
B.~Wong.
\newblock {ClosestNode. com: an open access, scalable, shared geocast service
  for distributed systems}.
\newblock {\em ACM SIGOPS Operating Systems Review}, pages 62--64, 2006.

\bibitem{Wong2005}
B.~Wong, A.~Slivkins, and E.~Sirer.
\newblock {Meridian: A lightweight network location service without virtual
  coordinates}.
\newblock In {\em ACM SIGCOMM Computer Communication Review}, volume~35, pages
  85--96. ACM, 2005.

\bibitem{Zhang2012}
Q.~Zhang, Q.~Zhu, M.~F. Zhani, and R.~Boutaba.
\newblock {Dynamic Service Placement in Geographically Distributed Clouds}.
\newblock In {\em Proceedings of the 2nd International Conference on
  Distributed Computing Systems}, pages 526--535. IEEE, jun 2012.

\end{thebibliography}

\begin{IEEEbiography}[{\includegraphics[width=1in,height=1.25in,clip,keepaspectratio]{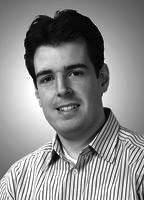}}]{Matthias Keller}
received his diploma degree in computer science from the University of Paderborn, Germany.
He is currently working as a research associate in Computer Networks group in University of Paderborn.
He focuses on adaptive resource allocation across 
wide area networks, designed a framework, and created a 
prototype testbed.
Previously, he worked at the Paderborn Centre for Parallel Computing as a research associate and at an international software development company as a software engineer.
\end{IEEEbiography}

\vspace*{-2\baselineskip}

\begin{IEEEbiography}[{\includegraphics[width=1in,height=1.25in,clip,keepaspectratio]{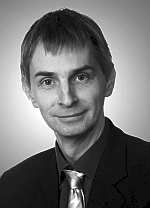}}]{Holger Karl}
received his PhD in 1999 from Humboldt University Berlin; afterwards he joined Technical University Berlin. Since 2004, he is Professor for Computer Networks at University Paderborn. He is also responsible for the Paderborn Centre for Parallel Computing and has been involved in various European and national research projects. His main research interests are wireless communication and architectures for the Future Internet.
\end{IEEEbiography}

\vfill

\end{document}